\documentclass[graybox]{svmult}

\usepackage{mathptmx}       
\usepackage{helvet}         
\usepackage{courier}        
\usepackage{type1cm}        
\usepackage{makeidx}         
\usepackage{graphicx}        
\usepackage{multicol}        
\usepackage[bottom]{footmisc}
\usepackage{amsmath}
\usepackage{color}
\usepackage{braket}
\usepackage{hyperref}
\definecolor{keywordcolor}{rgb}{0,0,1}

\makeindex             

\begin{document}

\title*{A highly efficient single photon-single quantum dot interface}
\author{Loic Lanco and Pascale Senellart}
\institute{Loic Lanco, CNRS-LPN Laboratoire de Photonique et de Nanostructures, Route de Nozay,  91460 Marcoussis, France; Université Paris Diderot. Paris 7, 75205 Paris CEDEX 13, France. \email{loic.lanco@lpn.cnrs.fr}\\
\and Pascale Senellart, CNRS-LPN Laboratoire de Photonique et de Nanostructures, Route de Nozay,  91460 Marcoussis, France; Physics Department, Ecole Polytechnique-RD128, 91763 Palaiseau CEDEX, France.  \email{pascale.senellart@lpn.cnrs.fr}}
%
%
\maketitle


\abstract{Semiconductor quantum dots are a promising system to build a solid state quantum network. A critical step in this area is to build an efficient interface between a stationary quantum bit and a flying one. In this chapter, we show how cavity quantum electrodynamics allows us to efficiently interface a single quantum dot with a propagating electromagnetic field. Beyond the well known Purcell factor, we discuss the various parameters that need to be optimized to build such an interface. We then review our recent progresses in terms of fabrication of bright sources of indistinguishable single photons, where a record brightness of 79\% is obtained as well as a high degree of indistinguishability of the emitted photons. Symmetrically, optical nonlinearities at the very few photon level are demonstrated, by sending few photon pulses at a quantum dot-cavity device operating in the strong coupling regime. Perspectives 	and future challenges are briefly discussed. \newline\indent
}

\section{Motivations}

To a large extent, semiconductor quantum dots (QDs) can be considered as artificial atoms. Strong confinement of the carriers in the three direction of space results in discrete energy levels and the coulomb interaction between carriers lead to a direct correspondence between the number of carriers in the QD and the energy levels \cite{marzin1994}. These properties make semiconductor QDs promising to implement quantum functionalities in a solid state system \cite{hidden}. Like real atoms, QDs can emit single photons \cite{michler2000} or entangled photon pairs \cite{akopian2006,NJPtoshiba}. The large oscillator strength of the transitions leads to a recombination time below one nanosecond allowing operation of the source in the GHz frequency range \cite{highfrequency, toshiba500mhz}. Finally, despite the coupling of the carriers to their vibrational and electrostatic environment, the emitted photons have been shown to present high degrees of indistinguishability, up to 96\% \cite{santoriindish,atesindish, indishentangled, mullerindish, heindish, gazzanonatcom}. \\
The anharmonicity of the energy levels in a QD also naturally opens the route toward single photon optical non-linearities \cite{Loo2012, Bose2012, vuckoprl2012, Reinhard2011}. The optical absorption of a photon resonant to a QD transition, leads to the creation of an electron hole pair which spectrally shifts the resonance for the absorption of a second photon. Such non-linearities could be used to implement optical quantum logic gates, with a gate operation speed determined by the QD  radiative transition rate \cite{waksgate}.\\

Finally, benefiting from the semiconductor technological possibilities, it is also possible to deterministically inject a carrier in a QD, using doped structures and electrical contacts \cite{electricQD}. The spin of such a carrier can be used as stationary quantum bit: while an electron spin presents coherence times in the few ns range \cite{electronspin1,electronspin2}, a hole spin can present a coherence time as long as 200 ns  \cite{holespin1,holespin2,holespin3,holespin4}. Since the main source of spin dephasing is the hyperfine interaction, spin echo techniques applied on the nuclear spin bath have allowed greatly increasing the electron spin coherence time \cite{electronspinecho}. Spin-orbit coupling in the excited charge state of the QD results in polarization selection rules for the optical transitions making it possible to optically manipulate and measure the spin. Applying a magnetic field in the Voigt configuration has also allowed full manipulation of the spin using virtual Raman optical transition \cite{spinraman} and very recently, spin photon entanglement has been reported \cite{spinphotonatac,spinphotonyoshie}. \\

All these properties have put the QD system in an interesting position to implement integrated quantum functionalities. To go beyond the demonstrations of principle, a major challenge is to make every functionality efficient. Indeed, QD based single photon sources present the attractive features of a solid state light source with true quantum statistics but suffer from low brightness, simply because total internal reflexion limits to a few percents the photons exiting the semiconductor. Techniques must be developed to collect every photon emitted by a QD. Symmetrically, optical quantum gates relying on the QD anharmonicity will only be demonstrated if one can ensure that every photon sent on a device will interact with the QD. Several approaches are pursued to build an efficient photon-QD interface \cite{highfrequency,claudon2010,reimer2012, gazzanonatcom,munsch2013,finleyprx}. In the last few years, the most successful ones have consisting in inserting the QD in a photonic structure, either a photonic wire \cite{claudon2010,reimer2012} or a microcavity \cite{gazzanonatcom}. The first approach relies on the single mode structure of a thin nanowire to guide the light emitted by the QD. This approach presents the advantage of offering a broadband high collection efficiency and could be applied to spectrally broad single photon emitters, like NV centers in diamond and colloidal QDs. In the case of QDs, the proximity  the surface has made the QD emission more sensitive to spectral diffusion phenomena \cite{munschapl2011} and dephasing of the carriers may be a limitation for obtaining indistinguishable photons. \\

Using cavity quantum electrodynamics  has been shown to be very efficient to build such an interface, and also to reduce the effect of dephasing induced by the solid state environment \cite{varoutsis2005, dousse2010,dousseapl2010}. When coupling a single QD to a confined optical mode, the light matter interaction is increased leading to an acceleration of spontaneous emission (Purcell effect) \cite{purcellgerard} or to new light-matter mixed states (strong coupling regime) \cite{SCRPBG,SCRpillar,SCRdisk}. Together with a full control on both the emission rate  and the radiation pattern of the QD emitted photons, this approach basically reduces the QD excited state lifetime hence its sensitivity to phonon assisted mechanisms \cite{varoutsis2005}, pure dephasing \cite{santoriindish} and spin-flip processes \cite{dousseapl2010}. First proposed in 1999 \cite{jlt2001}, the cavity based interface has faced many technological challenges regarding its implementation, because the QD must be precisely spatially and spectrally matched to the cavity mode, whereas QD mostly grow with random spatial and spectral properties. \\

This chapter reviews the recent progress made in this research line using a deterministic technique to couple a QD to a micro pillar cavity \cite{dousseprl2008}. In the first part, we discuss the physics of such a cavity based interface. While the Purcell factor and coupling to mode figures of merit are commonly discussed, we show that other parameters are critical for making the interface efficient.We also briefly describe the technology we have developed to have  a full control of the devices. In a second part, we review the progresses we have made in terms of fabrication of quantum light sources: brightness, indistinguishability of the emitted photons, purity of the single photon emission, electrically controlled sources. We also briefly present  a first application using a QD based bright source to implement an entangling quantum logic gate. In a third part, we present a study of the giant optical non-linearity for a QD-pillar device operating in the strong coupling regime. We further show that such a device allows monitoring single quantum events at the microsecond time scale.
In the last part, we discuss future challenges and objectives: spin-photon interface, scalability, limitations or possibilities provided by the solid state environment.

\section{Efficient quantum dot-photon interfacing}
\label{sec_efficient_interface}

\subsection{Basics of cavity-QED in a quantum dot-micropillar device}

\begin{figure}[b]
\centering
\includegraphics[width=12cm]{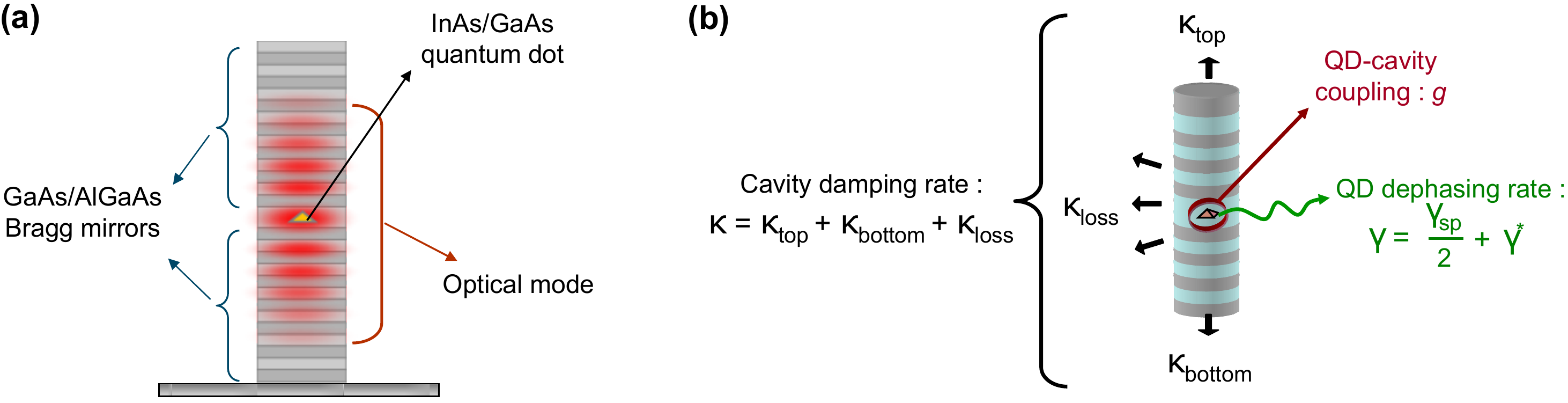}
\caption{\textbf{(a)} Typical structure of a quantum-dot/micropillar system. \textbf{(b)} Physical quantities describing the device behavior.}
\label{fig_sketch_device}
\end{figure}

Figure \ref{fig_sketch_device}a displays a sketch of a typical QD-pillar cavity system. Fabricating such a device requires, first, to fabricate a planar sample through molecular beam epitaxy: a layer of self-assembled InGaAs QDs is embedded into a GaAs cavity, sandwiched between two-distributed Bragg mirrors (alternating GaAs/AlGaAs layers). These Bragg mirrors induce the confinement of light in the vertical direction. Lateral confinement is then obtained by  etching a cylindrical micropillar, with a typical diameter of a few microns: a confined cavity mode is obtained with a cavity mode frequency $\omega_C$. In parallel, confinement of carriers in an InGaAs quantum dot leads to discrete energy levels, with a transition at frequency $\omega_{QD}$ between the QD ground state ($\ket{\mathrm{ground}}$) and its first excited state ($\ket{\mathrm{excited}}$). A maximal light-matter interaction is obtained when $\omega_{QD}\approx \omega_{C}$ (spectral matching), and when the InGaAs quantum dot is located at a maximum of the cavity mode intensity, i.e. at the center of the micropillar for the fundamental mode (spatial matching).\\

The important physical quantities governing the physics of such a cavity-QED device are sketched in Fig. \ref{fig_sketch_device}b:
\begin{itemize}
\item The \textbf{QD-cavity coupling strength $g$}: it describes the coherent interaction between the QD optical transition and the confined cavity mode. More precisely, it describes the rate at which a photon in the confined mode can be coherently converted into an electron-hole pair in the quantum dot, and vice-versa.
\item The \textbf{cavity damping rate $\kappa$}: it describes the incoherent dissipation associated to photons escaping the cavity. This damping rate is given by the sum of several contributions, through $\kappa=\kappa_{\mathrm{top}}+\kappa_{\mathrm{bottom}}+\kappa_{\mathrm{loss}}$. In the latter expression $\kappa_{\mathrm{top}}$ and $\kappa_{\mathrm{bottom}}$ are the damping rates associated to photons escaping through the top and bottom mirrors, while $\kappa_{\mathrm{loss}}$ is the damping rate associated to unwanted leakage through the unperfect micropillar sidewalls \footnote{We note that our cavity damping rate $\kappa$ is an intensity damping rate, whereas other references define $\kappa$ as a field damping rate: there is a factor 2 difference between these two possible definitions.}. 
\item The \textbf{QD decay rate $\gamma_{sp}$}: it describes the rate of the unwanted spontaneous emission of photons outside the cavity mode (as opposed to emission in the confined cavity mode, which is the desired emission channel).
\item The \textbf{QD pure dephasing rate $\gamma^*$}: it describes the rate at which the QD loses its coherence through pure dephasing processes. The total QD dephasing rate, denoted $\gamma$, is then the sum of a lifetime-limited contribution and of this pure dephasing contribution $\gamma^*$, through: $\gamma= \frac{\gamma_{sp}}{2}+\gamma^*$.\\
\end{itemize}

The objective for a QD-cavity device is to increase the strength of the coherent coupling $g$, as compared to the incoherent processes described by $\kappa$ and $\gamma$. In this respect, two regimes are usually introduced in cavity-QED:
\begin{itemize}
\item The \emph{strong-coupling regime} \cite{Andreani1999}, where $g$ is higher than both $\kappa/4$ and $\gamma/4$. In such a case, if the QD is in its excited state at a given time, it will be able to coherently emit a photon, absorb it, reemit it, reabsorb it, and so on, before dissipation occurs. In quantum words, the system experiences a Rabi oscillation at frequency $g$ between two states: 
$\ket{\mathrm{excited}} \otimes\ket{0 \ \mathrm{photon}}$  and $\ket{\mathrm{ground}}\otimes\ket{1 \ \mathrm{photon}}$.
\item The \emph{weak-coupling regime} \cite{purcellgerard}, where $g$ is smaller than either $\kappa/4$ or $\gamma/4$. In such a regime dissipative processes are faster than the coherent evolution, and therefore no Rabi oscillations can be observed. \\
\end{itemize}

Both the weak and the strong-coupling regimes provide a wide range of possibilities for quantum physics applications. For instance,  large values can be obtained for the QD emission rate in the confined mode, denoted $\Gamma$, which is given by $\Gamma=\frac{2g^2}{\kappa}$ in the weak-coupling regime \cite{Auffeves-Garnier2007}: this emission rate can be significantly higher than the emission rate  $\gamma_{sp}$ outside the cavity mode, ensuring the emission of easily collectable photons (see Section \ref{sec_bright_sources}). A well-known figure of merit in cavity-QED is thus the Purcell factor $F_p=\frac{\Gamma}{\gamma_{sp}}$ \cite{purcellgerard} \footnote{The Purcell Factor is usually defined as the ratio between the emission rate in the cavity mode, $\Gamma$, and the emission rate for a quantum dot in bulk GaAs, $\gamma_{bulk}$, but in a micropillar device $\gamma_{sp}$ is usually equal to $\gamma_{bulk}$.}.  Because $\gamma_{sp}$ is fixed by the properties of the QD material, optimizing the Purcell factor $F_p$ requires increasing the coupling strengh $g$ while at the same time decreasing the cavity damping rate $\kappa$. The optimization of $\kappa$ requires a significant number of GaAs/AlGaAs pairs in each Bragg mirror (typically more than 16 pairs in each mirror, to reach quality factors above a few thousands) and minimizing sidewall losses $\kappa_{loss}$, while the optimization of $g$ requires etching micropillars with small mode volumes and thus small diameters (typically less than a few microns) \cite{SCRpillar}. On top of that, the spectral matching condition ($\omega_{QD}\approx \omega_{C}$) and the spatial matching condition (QD at the micropillar center) also have to be fulfilled. The following section describes how both these requirements can be deterministically achieved with a specific in-situ lithography technique \cite{dousseprl2008}.

\subsection{Deterministic QD-cavity coupling through in-situ lithography}

Since 2005, many groups have worked on the deterministic coupling between a single QD and a cavity mode, using either top-down \cite{Badolato2005,dousseprl2008} or bottom-up approaches \cite{Gallo2008,Dalacu2010}. The first technological challenge, regarding this implementation, comes from the fact that QDs grow with random spatial locations on a planar surface (as is, for instance, the bottom mirror of a Bragg cavity). It is thus most probable, that, for a given quantum dot inside a randomly-etched micropillar, the QD location will not be at the maximum of the electromagnetic field. The second technological challenge comes from the wide inhomogeneous spreading of the QD transition frequencies $\omega_{QD}$, on a spectral range corresponding to a few tens of nanometers. In comparison, temperature adjustments allow tuning of the spectral mismatch $\omega_{QD} - \omega_{C}$ in a spectral range corresponding to approximately one nanometer only (for a typical temperature variation range between 4K and 50K). For a randomly-etched micropillar, the overall probability to find a spectrally matched QD at its center is thus of the order of $10^{-3}$.\\

\begin{figure}[b]
\centering
\includegraphics[width=12cm]{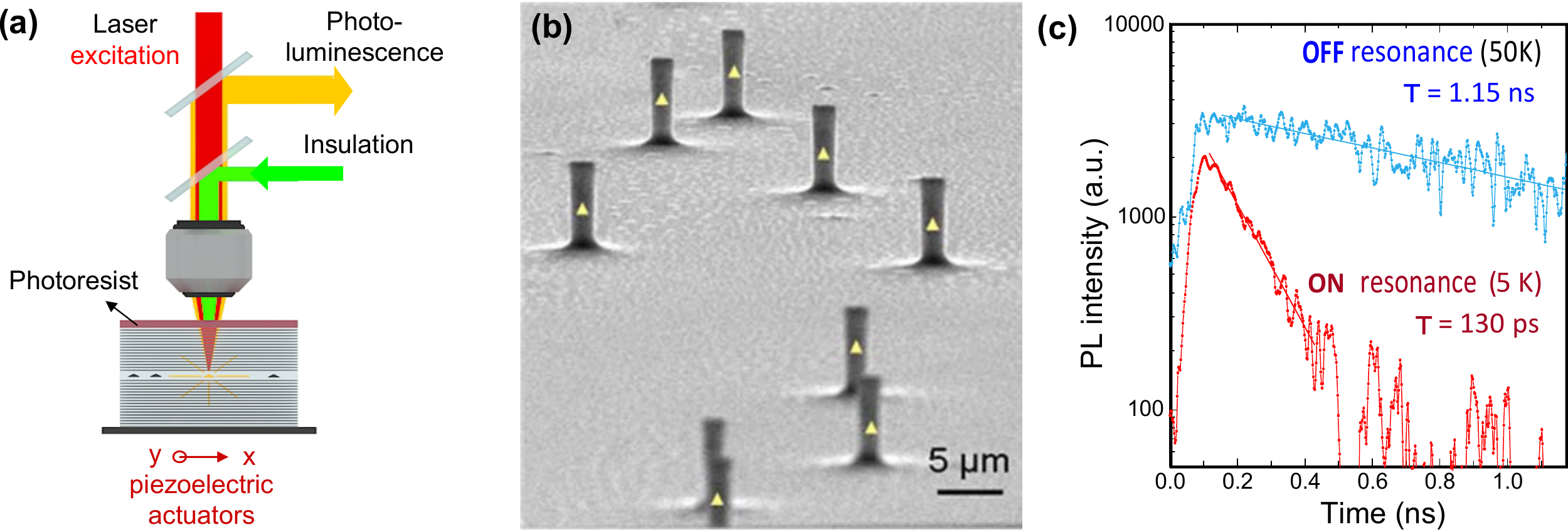}
\caption{\textbf{(a)} Principle of in-situ lithography. \textbf{(b)} SEM image of several deterministically-coupled micropillars. \textbf{(c)} Experimental demonstration of the Purcell effect with a deterministically-coupled pillar.}
\label{fig_in_situ}
\end{figure}

Standard dry etching of micropillars, starting from a planar Bragg cavity sample, requires a lithography step allowing to first define the positions and sizes of the micropillars. In the \emph{in-situ lithography} technique developped in 2008 \cite{dousseprl2008}, this step is performed inside a low-temperature cryostat. As sketched in Fig. \ref{fig_in_situ}a, the planar sample is  spin-coated with a positive photoresist and brought to low temperature; a 850 nm laser line is then used to excite the QD emission without exposing the resist. The emitted photoluminescence is analyzed with a spectrometer, allowing one to select a QD emission line and measure its intensity. Mapping this QD emission intensity as a function of the QD position, within the focused laser beam, allows measuring the QD position with 50 nm accuracy. A second laser, at 532 nm, spatially superimposed to the 850 nm one, is then used to expose a disk centered on the QD. Furthermore, the diameter of the exposed disk is adjusted in order to tune the micropillar diameter; this, in turn, allows tuning the pillar fundamental mode frequency $\omega_{C}$ and matching it to the QD emission frequency $\omega_{QD}$. The exposed disk is later used as a mask to etch the micropillar around the selected QD. This step is repeated as many times as desired for different QDs, so that one can fabricate many optimally coupled QD-pillar cavities on a single wafer, as illustrated in Fig. \ref{fig_in_situ}b.\\

A typical demonstration of the Purcell effect obtained with such devices is displayed in Fig. \ref{fig_in_situ}c. When the QD emission frequency $\omega_{QD}$ is tuned off resonance from the mode frequency $\omega_{C}$, emission in the confined mode remains negligible, and the QD emission lifetime is mainly governed by $\gamma_{sp}$. On the contrary, when $\omega_{QD}$ is tuned on resonance with $\omega_{C}$, emission in the confined mode becomes predominent: the QD emission lifetime is given by $\Gamma+\gamma_{sp}= (F_p+1) \gamma_{sp}$. From the lifetime measurements displayed in Fig. \ref{fig_in_situ}c a Purcell factor $F_p=7.8$ is deduced.

\subsection{Critical parameters: beyond the Purcell factor}

Controlling the QD emission rate is one thing; another is to take advantage of this control in order to develop a really efficient QD-photon interface. Such an interfacing must, ideally, go both ways: transfer of information from an incident photon to a QD, and from a QD to an extracted photon. The perspectives offered by an efficient QD-photon interface are very wide, as will also be discussed in the last section of this chapter (Sec. \ref{sec_future_challenges}). In the following we will mainly focus on two major aspects of QD-photon interfacing: the development of ultrabright sources of indistinguishable single photons \cite{gazzanonatcom} (Sec.  \ref{sec_bright_sources}) and the demonstration of an optical nonlinearity with few-photon pulses \cite{Loo2012} (Sec.  \ref{sec_few_photon_nonlin}). In the former application, the QD-pillar device is used to \emph{emit} single photons with specific properties. In the latter, it is used to \emph{receive} incident photons, and subsequently transmit or reflect them, depending on the QD state. Here we discuss the critical parameters which characterize the quality of a QD-pillar device for both applications.\\

As regards photon emission, quantum communication applications ideally require deterministically-triggered emission of indistiguishable single photons. One thus has to control at the same time:
\begin{itemize}
\item The \emph{fraction of photons emitted in the mode}, denoted $\beta$. Indeed, only photons emitted in the confined mode can be efficiently collected through the cavity top mirror. $\Gamma$ being the emission rate in the mode, and $\gamma_{sp}$ the emission rate outside this mode, the fraction $\beta$ is given by $\beta=\frac{\Gamma}{\Gamma+\gamma_{sp}}$, i.e. $\beta=\frac{F_p}{F_p+1}$ with $F_p$ the Purcell factor defined above. A large Purcell factor is required to obtain $\Gamma\gg\gamma_{sp}$, i.e. emission in the mode with a fraction $\beta$ close to unity.
\item The \emph{single-photon wavepacket indistinguishability}, usually denoted $\frac{T_2}{2 T_1}$. This is a figure of merit indicating if the single-photon wavepacket is close to the Fourier-transform limit, where the photon coherence time $T_2$ equals twice its lifetime $T_1$. In our case $T_1^{-1}=\Gamma+\gamma_{sp}$ (sum of the emission rates into and outside the confined mode), while $T_2^{-1}=\frac{\Gamma+\gamma_{sp}}{2}+\gamma^*$ also includes the pure dephasing described by $\gamma^*$. A large Purcell factor is usually required to obtain $\frac{\Gamma}{2}\gg\gamma^*$, i.e. neglible dephasing and thus a wavepacket close to the Fourier-transform ideal limit. \\
\end{itemize}

These two separate conditions $\Gamma\gg\gamma_{sp}$ and $\frac{\Gamma}{2}\gg\gamma^*$ can be fulfilled both at the same time if $\frac{\Gamma}{2}\gg\gamma$, where $\gamma$ is the total QD dephasing time previously defined: $\gamma=\frac{\gamma_{sp}}{2}+\gamma^*$. Because $\frac{\Gamma}{2}=\frac{g^2}{\kappa}$, this allows introducing a fundamental quantity which is the device \textbf{cooperativity}, denoted $C$:
\begin{equation} \label{eq_coop}
C=\frac{g^2}{\kappa \gamma}
\end{equation}
This cooperativity is a well-known figure of merit in cavity-QED, first introduced with cold atoms \cite{Turchette1995}. It compares the strength of the coherent interaction (governed by $g$) to that of the incoherent processes (governed by $\kappa$ and $\gamma$), and indicates how strongly the presence of the QD transition modifies the optical properties of the device.\\

We point out that different notations are sometimes adopted in the literature. As an example, in atom cavity-QED the pure dephasing term $\gamma^*$ can be considered equal to zero, as is in Chapter 1 of this book. In such a case the spontaneous emission rate $\gamma_{sp}$ is equal to twice the total dephasing rate, and the Purcell Factor (denoted $f$ in Chapter 1 and $F_p$ here) is equal to twice the cooperativity. We also note that with our definition, the cavity damping rate $\kappa$ is an intensity damping rate, whereas it is a field damping rate in Chapter 1: the cavity linewidth is thus equal to $\kappa$ in the present chapter, but to $2 \kappa$ in Chapter 1.\\

Another crucial quantity to be optimized is the \textbf{top-mirror output-coupling efficiency}, denoted $\eta_{\mathrm{top}}$:
\begin{equation}
\eta_{\mathrm{top}}=\frac{\kappa_{\mathrm{top}}}{\kappa}
\end{equation}
This top-mirror output coupling efficiency gives the fraction of photons from the confined cavity mode that escape through the top mirror. This derives from the fact that the cavity damping rate is the sum of contributions from several channels: $\kappa=\kappa_{\mathrm{top}}+\kappa_{\mathrm{bottom}}+\kappa_{\mathrm{loss}}$, so that $\frac{\kappa_{\mathrm{top}}}{\kappa}$ measures the probability for escaping through the top-mirror channel. For single-photon emission applications, it is crucial to approach $\eta_{\mathrm{top}}\approx 1$ in order to collect efficiently photons from the cavity mode: this  requires an asymetric design with a highly reflective bottom mirror (so that $\kappa_{\mathrm{top}}\gg\kappa_{\mathrm{bottom}}$), as well as low sidewall losses (so that $\kappa_{\mathrm{top}}\gg\kappa_{\mathrm{loss}}$). As will be discussed in Section \ref{sec_few_photon_nonlin}, $\eta_{\mathrm{top}}$ also plays a crucial role in resonant excitation experiments where photons are received and then reflected or transmitted by the device.\\

Finally, regarding photon reception experiments, one must not forget the requirement that photons have to be injected efficiently into the fundamental cavity mode: to do so, one has to optimize the spatial overlap between the free space optical beam and the confined mode (exactly as one would do to efficienctly inject light into an optical fiber). The overlap integral between these two spatial shapes gives us another figure of merit, the \emph{input-coupling efficiency} of our experiment, denoted $\eta_{\mathrm{in}}$. Contrary to $C$ and $\eta_{\mathrm{top}}$, which are related to the quality of the device technology, $\eta_{\mathrm{in}}$ is governed by the experiment and can be optimized with a careful optical alignment. Because the fundamental mode of the pillar cavity present a high overlap with a gaussian mode,  $\eta_{\mathrm{in}}$ values close to unity can be obtained \cite{Loo2010,Arnold2012}.

\section{Ultrabright single photon sources}\label{sec_bright_sources}

\subsection{Why are bright single photon sources  desirable?}
Although single QDs have been shown to emit  single photons as early as  fourteen years ago \cite{michler2000}, most optical quantum communication and quantum computation protocols are still mostly implemented using parametric down conversion (PDC) sources. The main reason for this is that PDC sources present the main advantage of generating highly indistinguishable photons at room temperature. Their main limitation is their photon statistic, which is at best Poissonian (for non heralded sources) and which strongly limits the operation rate of the source in order to minimize multi-photon events. Over the years, multi-photon events have been dealt with error correction protocols and the number of entangled photon has recently reached a record value of eight \cite{8photon}. Yet,  the low brightness and the multi photon events of PDC sources may soon put a strong barrier to the scalability of photonics quantum networks, simply because of the  measurement time exponentially increases with the number of photons. 

A QD based single photon source, even highly indistinguishable, is not of much interest if one cannot collect more than few percents of the emitted photons, 5-10 \% being the typical operation rate of non-heralded PDC sources.
On the contrary, a very bright QD  source of highly indistinguishable photon could have strong potential in this context. Recent progresses in the community  indicate that such a source is within reach. We now present the recent progresses we made in term of QD based single photon sources for quantum information processing by inserting QDs in micro pillar cavities.

\subsection{Demonstration of single photon sources with record brightness}
\begin{figure}[h!]\begin{center}\includegraphics[width=0.9\linewidth]{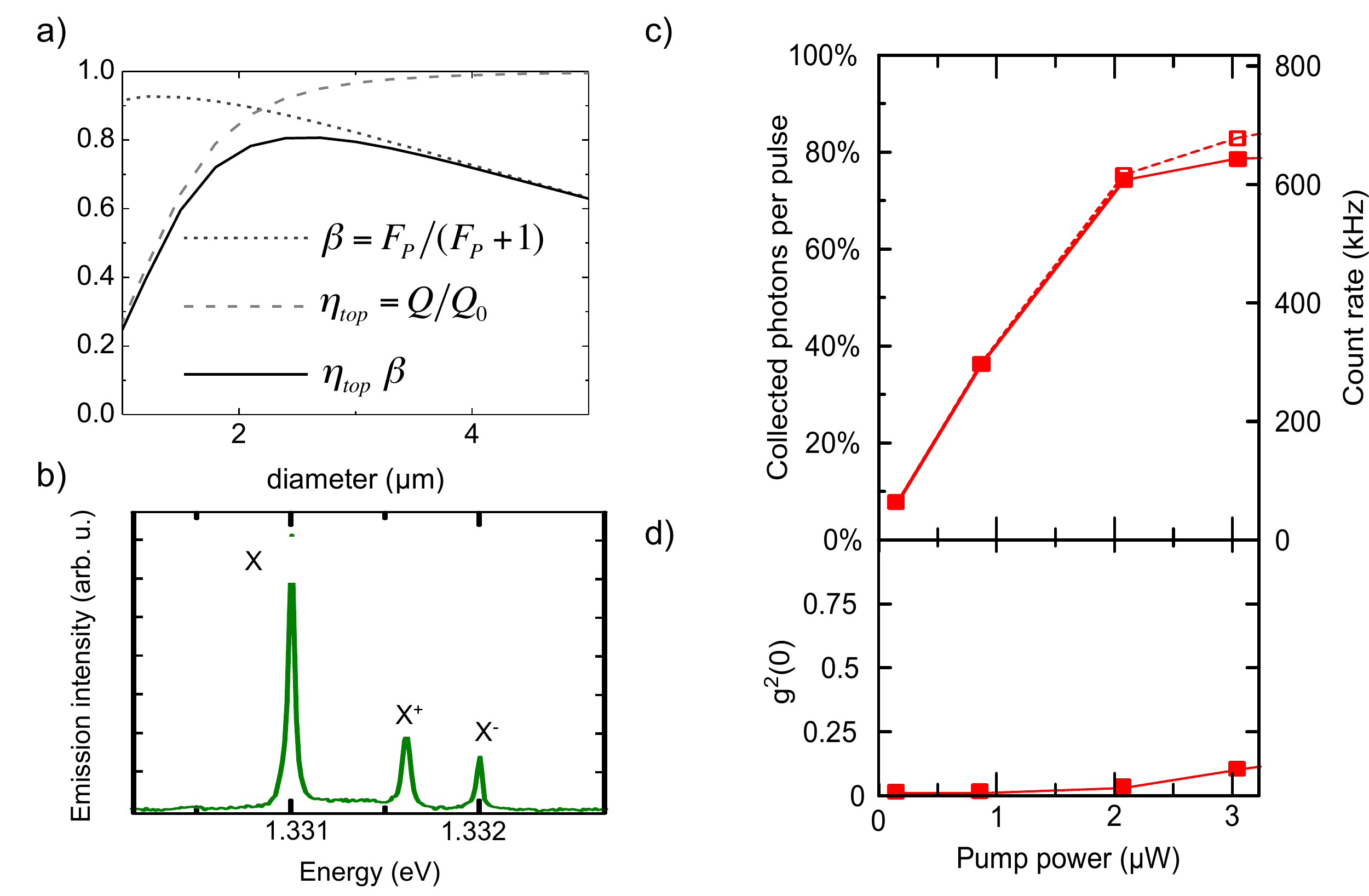}\caption{ {\bf Ultrabright single photon source} a: Dashed line: $\eta_{top}=\frac{Q}{Q_0}$  as a function of the pillar diameter. Dotted line: coupling to the mode $\beta=\frac{F_P}{F_P+1}$. Solid line: extraction efficiency  $\eta_{top} \beta$. b: Example of a QD spectrum  under non-resonant excitation where several charged exciton emission lines are observed. c: Count rate on the detector (right axis) and brightness (left axis) as a function of the excitation power. Solid line correspond to raw data ($\mathcal{B}$), dashed line is corrected from multi photon emission ($\mathcal{B}_{corr}=\mathcal{B}\sqrt{1-g^{(2)}(0)}$). d: $g^{(2)}(0)$ as a function of the incident power. The multiphoton emission is very small at low powers, and slightly increases when higher powers are used.}\label{bright}\end{center}\end{figure}

We define the brightness of the source as the number of collected photon per excitation pulse in the first collection lens. For a high excitation power, one can assume that at least an electron hole pair is created in the QD. The QD high quantum efficiency means that this electron hole pair will radiatively recombine with a probability close to one \cite{bleuse2011,gazzano2011}. This first step describes the photon creation efficiency. To obtain a bright source, high creation efficiency must be combined with high collection efficiency. As explained in section \ref{sec_efficient_interface}, in the weak coupling regime, the collection efficiency is given by the coupling to the mode $\beta=\frac{F_P}{F_P+1}$ multiplied by the out coupling efficiency $\eta_{top}$. To collect all the emission from one side of the pillar, we use an highly asymmetric cavity, where the transmission of the top mirror strongly exceeds the one of the bottom one $\kappa_{\mathrm{top}}\gg\kappa_{\mathrm{bottom}}$. In this case, $\eta_{top}$ is only limited by the side losses and is given by $\eta_{top}=\frac{\kappa_{\mathrm{top}}}{\kappa}=\frac{Q}{Q_0}$ where $Q$ and $Q_0$ are the quality factor of the pillar and planar cavities.  The dashed line in figure \ref{bright}a show this ratio for  a typical etching process, starting from a planar cavity with $Q_0=3000$. During the pillar etching process, some roughness can develop on the pillar sidewalls, resulting in a decreasing $Q$ when decreasing the pillar diameter. The  corresponding $\beta$ (dotted line) increases as the mode volume decreases. As a result, the collection efficiency $\eta_{top} \beta$ presents an optimum around 80\% obtained for a pillar diameter around 2 to 3 $\mu m$ here. \\

To reach such a high value, we need to consider the actual typical spectrum of a QD under  non-resonant excitation (see Figure \ref{bright}b ). The QD emission spectrum consists in discrete emission lines, each corresponding to a well defined charge  state of the QD. Although the samples are not intentionally doped, several emission lines can be observed, the neutral exciton (X), the positively charged exciton (X$^+$) and the negatively charged exciton (X$^-$). The observation of these three lines shows that depending on the excitation cycles, the QD will be in either one of these states.  To reach a creation efficiency close to one, ideally, the QD should be in only one of these states with a high probability. Gated structures can be used to control this charge state \cite{electricQD}. In this first demonstration, no electrical control of the source is used. We use the possibility to select the QDs presenting only a single emission line during the in-situ lithography process.\\

 Several dozens of sources are therefore fabricated selecting QDs with a bright single emission line and spectrally matched to pillar cavity mode with diameter around 2-3 $\mu m$. The source brightness is measured using a simple experimental setup consisting of a collection objective, mirrors and cubes and a spectrometer coupled to an avalanche photo diode (APD). Each optical component transmission or detection efficiency has been carefully measured using an attenuated pulsed laser at the QD emission wavelength.  Figure \ref{bright}c presents the number of counts measured on the APD as a function of a pulsed excitation power at 82 MHz. At high power, when the probability to create at least an electron hole pair is close to one, a $0.65$ MHz count rate is measured on the detector. Taking into account the overall setup efficiency of $0.97 \%$, this corresponds to a brightness around $\mathcal{B}=83\pm 8 \%$. The purity of the single photon source is also measured through photon correlation measurements (figure \ref{bright}d). A very low $g^{(2)}(0)$ below 0.05 is observed up saturation. The corrected brightness $\mathcal{B}_{corr}=\mathcal{B}\sqrt{1-g^{(2)}(0)}$ amounts to $78\pm 8 \%$, a record value for a single photon source. Brightness ranging between $60$\% and $79$\% have been obtained in this first generation sample, with $\eta_{top} \approx 1$ and $\mathcal{B}  \approx \beta$, with Purcell factors ranging in the $2$ to $3.5$ range. In a new generation of sample, similar values have been obtained using an adiabatic design for the cavity as proposed in \cite{adia}. Such a design reduces the effect of sidewall losses, allowing to maintain higher quality factors for smaller pillar diameters. Brightness in the $75 \pm 7 \%$ range are obtained with $\beta \approx 1$ and $B \approx \eta_{top}$, with Purcell factors around 10. In between these two regimes, i.e. with slightly smaller $F_P$, brightnesses in the $90\%$ range should be reached. 

\subsection{Purity of the single photon emission}

While the QD emission usually presents a quantum statistic with $g^{(2)}(0)<0.5$, the observed values for the second order autocorrelation function $g^{(2)}(0)$ can significantly vary from one device to another and from one measurement to another on the same device. In the literature, two phenomena are mainly proposed to explain the residual $g^{(2)}(0)$: multiple capture processes \cite{peter2007} and cavity feeding effects\cite{feeding1,feedingjan,feeding2}. We now discuss these various phenomena and show that only recapture processes affect the single photon purity for deterministically coupled devices and how one can systematically obtain a nice single photon purity with the appropriate excitation conditions.\\

\begin{figure}[h!]\begin{center}\includegraphics[width=0.8\linewidth]{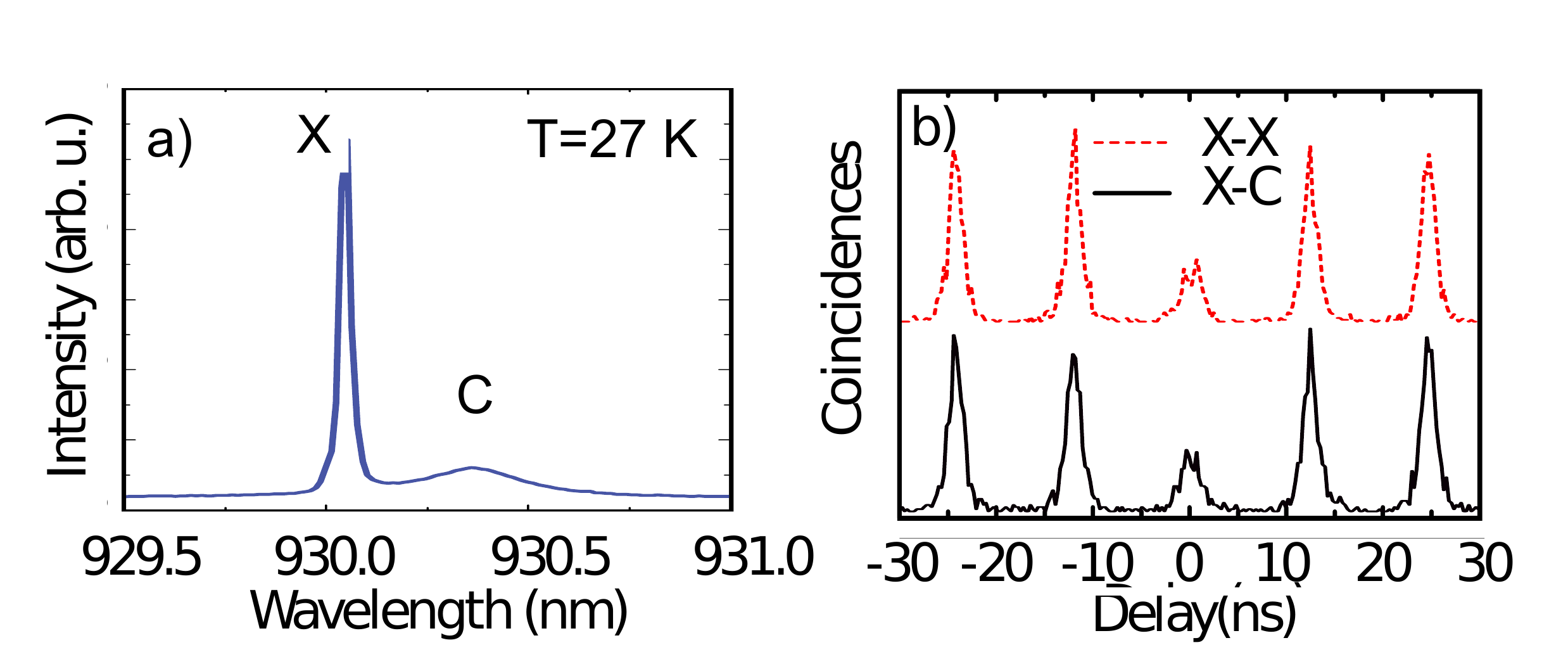}\caption{  {\bf Origin of the cavity emission line} a : Spectrum of a deterministically coupled QD-pillar device for a non-zero detuning between the QD exciton line and the cavity resonance. Two emission lines are observed : the D exciton line and a emission line close to the cavity resonance. b : Photon correlation measurements. Red : auto-correlation of the exciton line. Black : cross correlation between the exciton and cavity line.}\label{feeding}\end{center}\end{figure}

When increasing the excitation power, some emission background is sometimes observed  together with the   discrete emission lines of the QD \cite{selftuned}. Because of the strong phonon and coulomb interaction with their solid state environment, few percent of QD emission is emitted on a broad spectral range.  When the QD is in a cavity, this broad emission emission is enhanced by the cavity resonance,  leading to the so called cavity feeding effect, namely the observation of an emission at the cavity resonance, even when no QD optical transition is resonant to the cavity. 
 When several QDs are inserted in the device, the emission at the cavity mode energy can arise from several spectrally non-resonant QDs. Such emission at the cavity energy can significantly decrease the single photon purity. 
 
 However, we show that when a single QD is coupled to the cavity line, cavity feeding effects  do not explain a bad single photon purity. The emission spectrum of a deterministically coupled QD device when the QD resonance is not matched to the cavity line is presented in figure \ref{feeding}.a. Two emission lines are observed, one corresponding to the QD resonance, the other close to the cavity resonance. Figure \ref{feeding}.b. presents the measured auto correlation  function of the exciton line where $g_{X,X}^{(2)}(0)\approx 0.4$. The same value is observed for the cross correlation between the exciton and the cavity line $g_{C,X}^{(2)}(0)=g_{X,X}^{(2)}(0)$. If the cavity mode arised from several QD emission lines then $g_{C,X}^{(2)}(0)> g_{X,X}^{(2)}(0)$. This observation shows that the cavity like emission arises from the very same QD line and can be accounted for by the phonon sidebands. It cannot explain  the bad  single photon purity illustrated here.\\

\begin{figure}[h!]\begin{center}\includegraphics[width=0.8\linewidth]{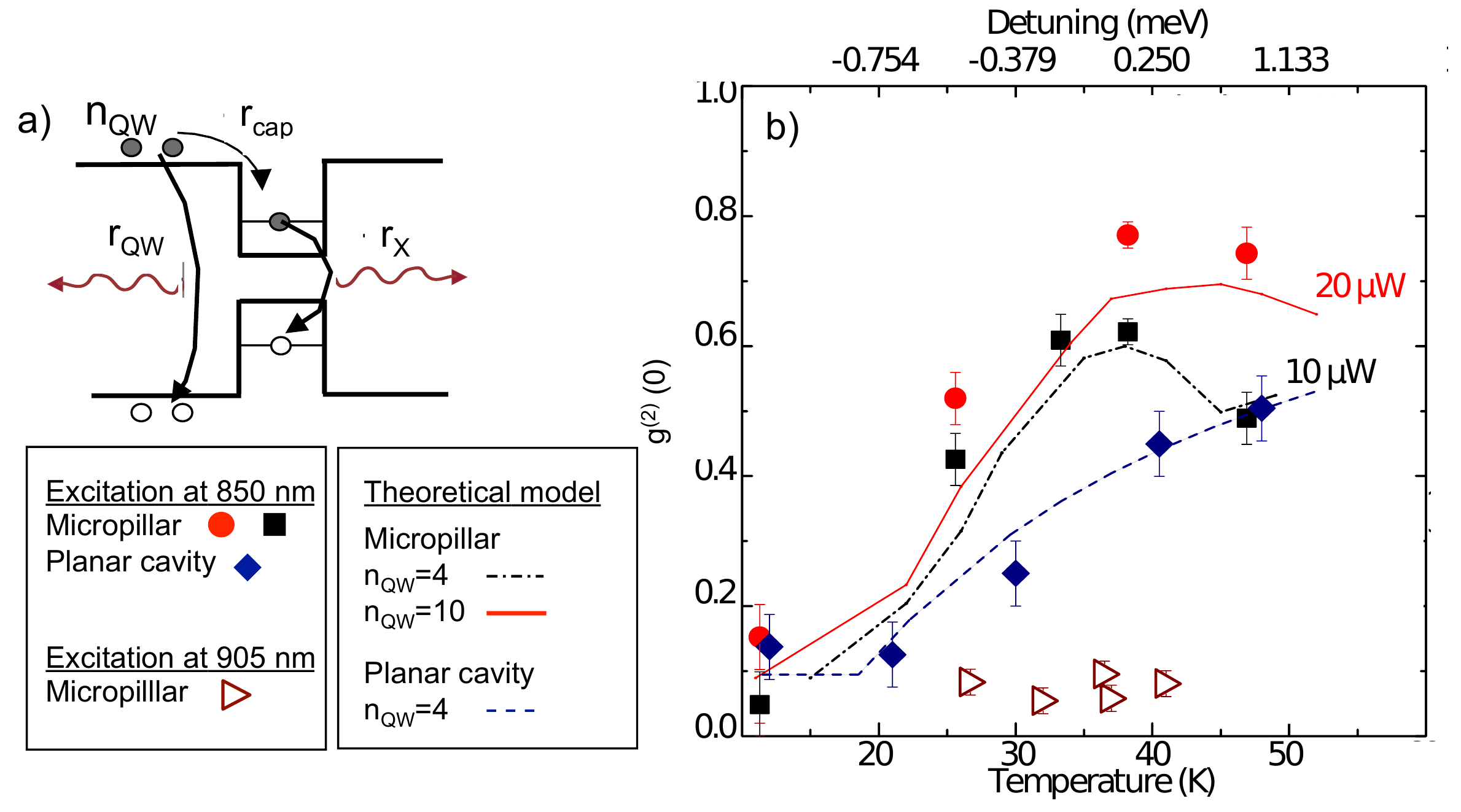}\caption{{\bf Influence of multiple capture processes  on the single photon emission.} a :  Schematics of the processes involved in the QD single photon emission under on resonant excitation (see text). b: Exciton auto-correlation function under non-resonant excitation as a function of the temperature for a QD in a pillar cavity (circles and squares) and planar cavity (diamonds). Symbols are experimental data, lines theoretical predictions detailed in \cite{valerianAPL}. Triangles: Exciton auto-correlation function under quasi-resonant excitation. For the QD in the pillar device, the detuning from the cavity mode is indicated on the top scale.   }\label{capture}\end{center}\end{figure}

To explain the bad single photon purity presented here, figure \ref{capture}  recalls the main mechanisms involved in the single photon generation for a QD system pumped non-resonantly. A pulsed non-resonant excitation creates a population of carriers $n_{QW}$ in the wetting layer or GaAs barriers. These carriers  recombine radiatively or non-radiatively with a rate $r_{QW}$ or are captured in the QD with a rate $r_{cap}$. Assuming that there is only a single confined exciton state in the QD, the QD exciton and biexciton states radiatively recombine with rates $r_X$ and $r_{XX}$. The guarantee for a good single photon source is that when the QD exciton recombines, there are no carriers left in the barriers that can be captured in the QD, namely that $r_X, r_{XX} \ll r_{QW}, r_{cap}$. As a result, several mechanisms can degrade the single photon purity. 

On one hand, a very high quality barrier where carrier can spatially diffuse on long time and spatial scales would decrease $r_{QW},r_{cap}$. Increasing temperature can also increase the lifetime of the carriers in the barrier.
On the other hand, shortening the exciton radiative lifetime should also reduce the single photon purity under non-resonant pumping. This is what is evidenced in figure \ref{capture}. $g_{X,X}^{(2)}(0)$ is plotted as a function of temperature (bottom scale) corresponding to a detuning with the cavity mode (top scale). These measurements are taken  for an excitation power close to saturation. High value for $g_{X,X}^{(2)}(0)$ are observed for the whole temperature range. As a reference, the $g_{X,X}^{(2)}(0)$ for a QD in the planar structure (not experiencing Purcell effect) is shown; a continuous degradation of the single photon purity is observed increasing temperature, because of a decreased of $r_{QW}, r_{cap}$. For the QD in the cavity, the single photon purity is further degraded when the QD is brought in resonance with the cavity mode, increasing $r_X$.\\ 

Because the single photon purity is degraded by multiple capture mechanisms, a good single photon purity can be obtained by a direct creation of the carriers inside the QD: the relaxation of carriers between confined energy levels is very efficient and hardly temperature dependent. This is what is demonstrated with the open symbols in figure \ref{capture}.b: a very good single photon purity, with $g_{X,X}^{(2)}(0)<0.08$ is observed on the whole temperature and detuning range.

\subsection{High indistinguishability through a  control of the QD environment }

Chapter 13 of this book recalls the requirements for obtaining indistinguishable photons, namely the photons should be identical in polarization, energy, spatial and temporal mode. Finally, the most demanding requirement concerns the emission of single photons with a Fourier transform limited spectrum.  This last requirement immediately brings the question of environment induced dephasing for an emitter in a solid state system. Several mechanisms can limit the photon indistinguishability. Coupling with acoustic phonons leads to the appearance of phonon sideband emission \cite{besombes2002,favero2003,peter2004}, while coupling with optical phonons induces with pure dephasing of the zero phonon line. Moreover, charges in the QD surrounding (either fluctuating charges in traps or optically created charges in the barrier) create a fluctuating electric field, leading to a Stark induced fluctuation of the emission energy. Depending on the relative time scale between the charge fluctuations and the exciton radiative recombination, this charge noise will result in a homogeneous broadening (pure dephasing) or an inhomogeneous one (spectral diffusion) \cite{berthelot}. Finally, obtaining indistinguishable photons also  depends on the dynamics of carrier relaxation and emission in the system. Very high pumping, which creates many electron hole pairs in the QD, delay the emission of the exciton \cite{senellarthours2005} and lead to a strong jitter in the QD emission dynamics. 

Despite these possible limitations, QDs have been shown to emit indistinguishable photons as early as 2002 \cite{santoriindish}, with mean wave packet overlap as large as $80\%$. Since then, many works have reported on the emission of indistinguishable photons \cite{atesindish,indishentangled,mullerindish,heindish}. In most works, the indistinguishability is  below $80\%$ and the origin of this limitation is not clear. Very recently, pure resonant excitation has allowed the observation of indistinguishability of $96 \%$ \cite{heindish} , bringing the QD source close to the quality of PDC sources. Yet, this was obtained for a low source brightness.\\

\begin{figure}[h!]\begin{center}\includegraphics[width=0.65\linewidth]{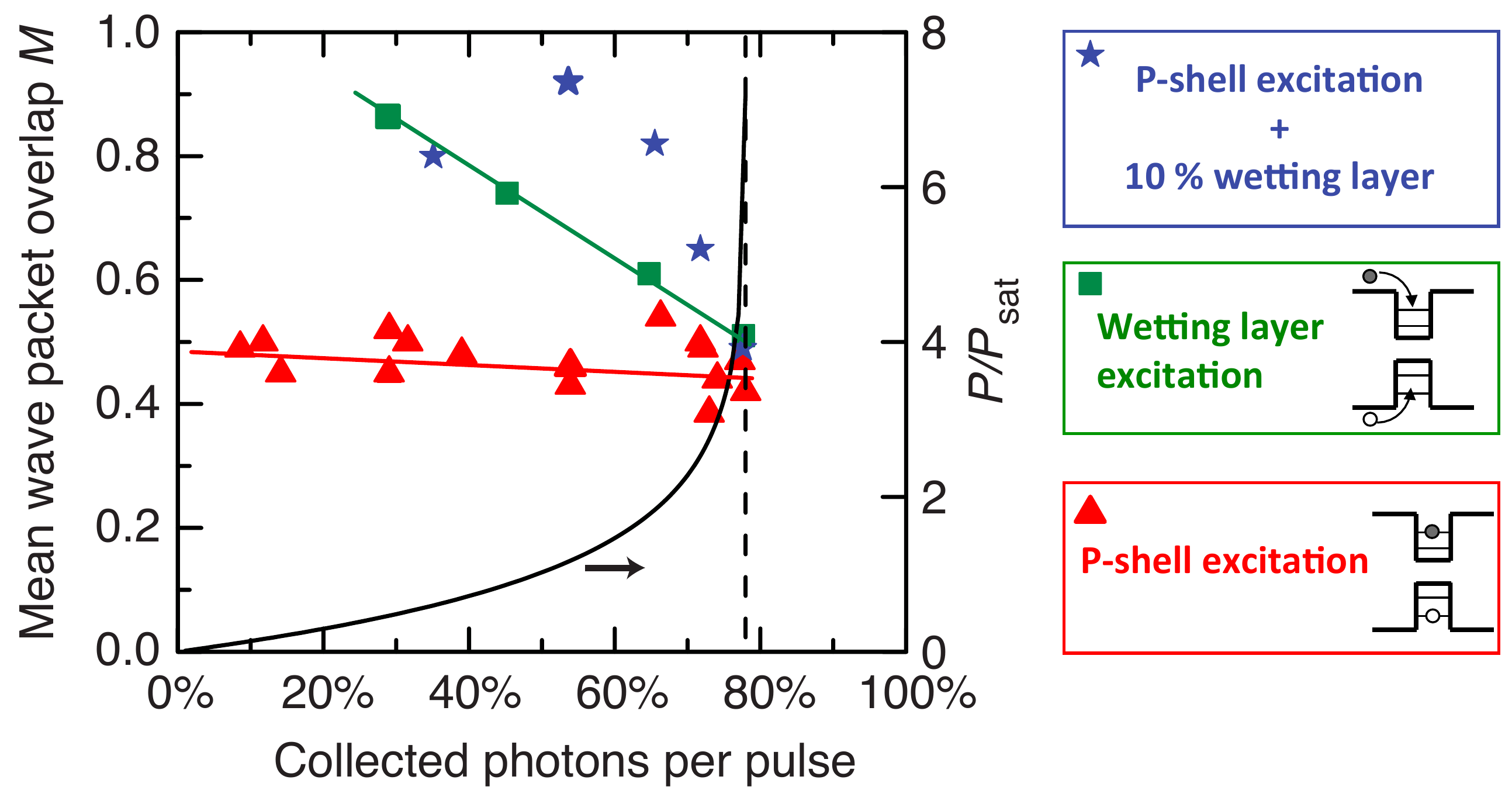}\caption{ {\bf Brightness and indistinguishability.} Mean wave packet overlap characterizing the indistinguishability of successively emitted photons as a function of the source brightness. The right scale shows the excitation power normalized to the saturation excitation power corresponding to a given brightness. Triangles: quasi resonant excitation in the QD. Squares: excitation in the wetting layer. Blue: two color excitation scheme: quasi resonant excitation (90\% of the time integrated count rate)  combined with wetting layer continuous wave excitation (10 \% of the time integrated count rate).   }\label{indish}\end{center}\end{figure}

 We have studied the indistinguishability of a QD-pillar single photon source as a function of the source brightness. When creating the carriers in the surrounding barriers (figure \ref{indish}, green symbols), a high photon indistinguishability, characterized by a mean wavepacket overlap M=0.82, is observed at a source brightness of 30\%.  When increasing the source brightness, M continuously decreases: additional carriers optically created in the QD surrounding create a fluctuating electrostatic environment. To circumvent this effect, carriers are directly created in the excited state of the QD (red symbols). Surprisingly, the source indistinguishability is even lower, independently of the source brightness. Considering these two sets of measurements, we deduce that under low power non-resonant excitation, non-resonantly created carriers fill deep traps around the QD and stabilize its electrostatic environment.   To combine high brightness with high indistinguishability, we have therefore used a two color excitation scheme (blue symbols): a strong pumping directly creating excitons into an excited QD state together with a weak non-resonant pumping to fill traps. Doing so, we demonstrate a mean wavepacket overlap as high as 92\% (82\%)  for a source brightness of 53 \% (65\%). These values are close to the best values ever reported on QD system, combined here with a high brightness.\\
 
 \begin{figure}[h!]\begin{center}\includegraphics[width=0.65\linewidth]{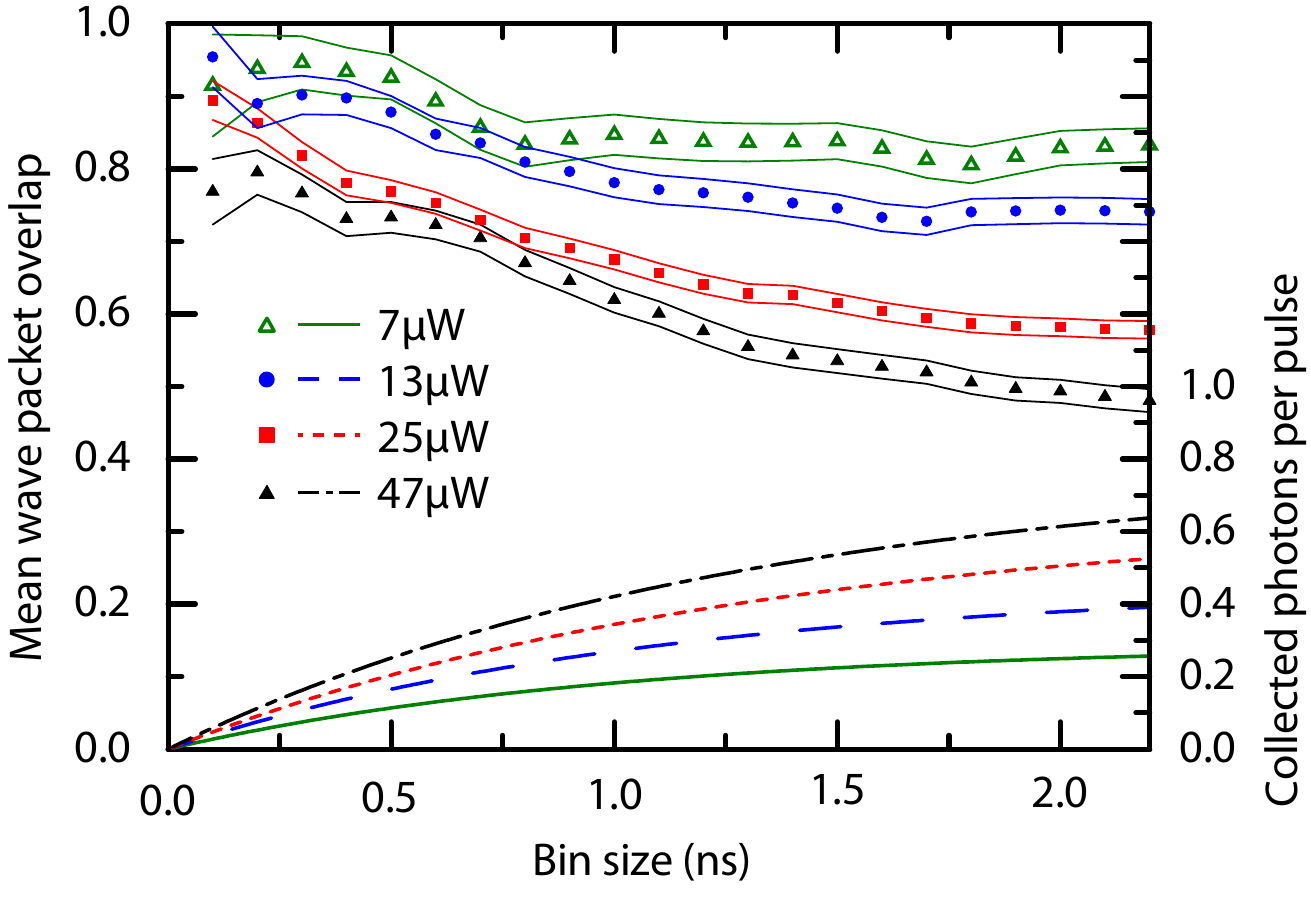}\caption{ {\bf Brightness and post temporal selection.}  Measured  photon mean wave packet overlap as function of the time bin size of detection for three excitation powers in the wetting layer. The lines indicate the experimental error bars. The brightness corresponding to the time bin size is shown on the right scale.}\label{indishbin}\end{center}\end{figure}

Finally, we analyze the dynamics of the indistinguishability by performing a temporal post selection of the emitted photons. Figure \ref{indishbin} presents the indistinguishability of the source as a function of the time bin for the analysis. This temporal post selection reduces the brightness of the source as indicated on the right axis. The measurements presented here correspond to an excitation in the wetting layer. For all excitation powers, a higher indistinguishability is observed at shorter time delay: the earlier the photon is emitted after the excitation pulse, the less the exciton has experienced dephasing.

\subsection{Electrically controlled sources}

Inserting the QD in a doped structure and applying an electric field is a very efficient tool in the context of building a solid state quantum network. It first allows deterministically injecting an electron or hole in the QD \cite{electricQD} in order to build a spin based quantum memory. It has also been used to control the coupling between the two linearly polarized exciton states and produce entangled photon pairs \cite{bennett2010} (see chapter 10). Applying an electric field can allow tuning the QD emission energy through the Stark effect, an interesting property to implement quantum interferences between two sources \cite{patel2010}. Finally, a doped structure and an applied bias around the QD layer helps stabilizing the QD charge environment and reduce charge induced dephasing.\\ 

\begin{figure}[h!]\begin{center}\includegraphics[width=0.9\linewidth]{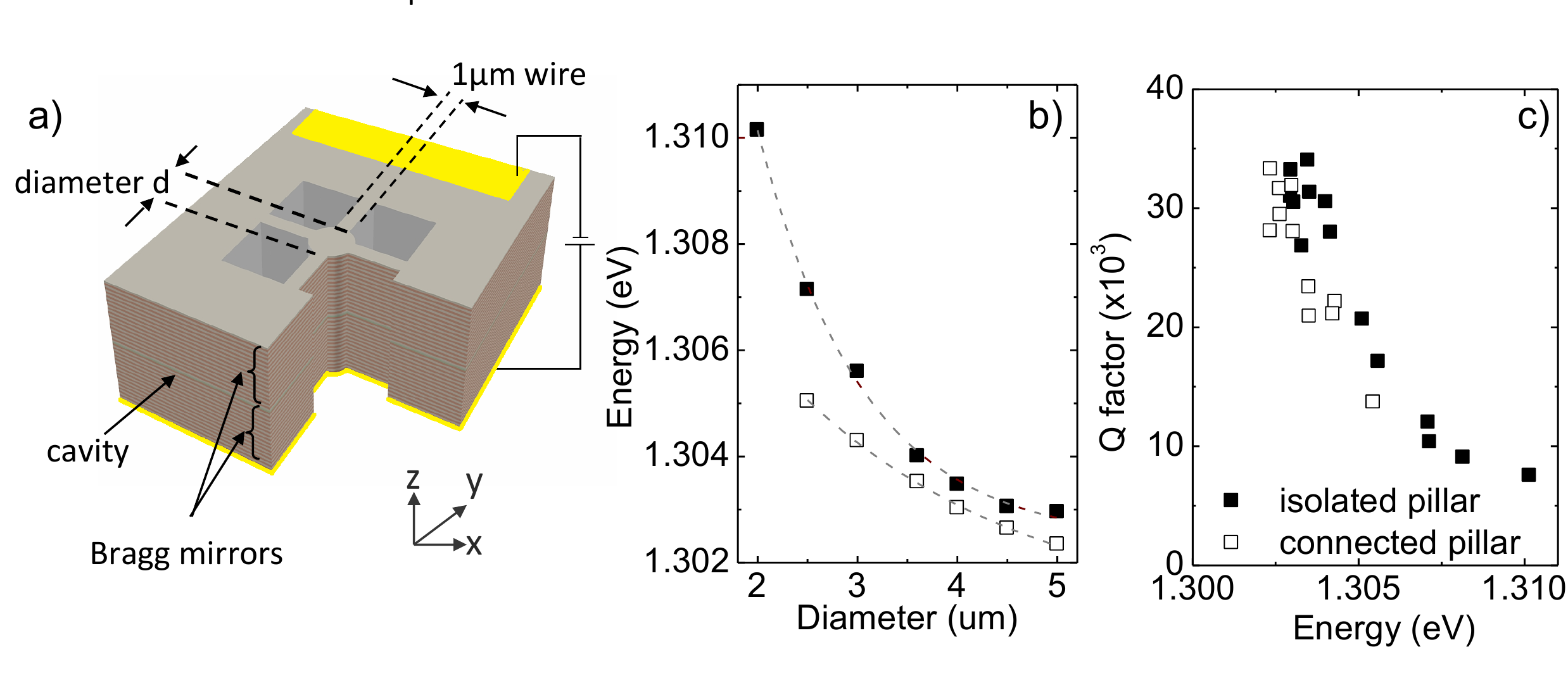}\caption{ {\bf Connected pillar cavity.} a: Schematic of the connected pillar structure used to implement electrical contacts on a  cavity. b: Energy of the connected micropillar (open) and isolated (solid) pillar fundamental mode as a function of pillar diameter. c: Quality factor of the fundamental mode a function of the mode energy for isolated (solid) and connected (open) pillars.}\label{connected1}\end{center}\end{figure}

Combining an electrical control with a good extraction efficiency is technologically challenging. Pioneer works have developed a technology consisting in planarizing a micropillar sample and defining an anular contact on top of a micropillar \cite{contactpillar}. Another approach has consisted in used oxide aperture cavities \cite{highfrequency}. In such structures, the carrier injection is very close to the quantum dot layer, a favorable approach to obtain fast operation of the electical control. On the other hand, a precise control over the oxidation process is needed to control the cavity energy. \\

We have proposed another approach to obtain an electrical control of a QD in a cavity \cite{anna2014}. The cavity consists in a micropillar, connected with one dimensional wires to a larger frame, where the electrical contact is defined (figure \ref{connected1}.a.). To study the optical properties of the connected pillar cavity,  a preliminary study  was conducted on a high quality factor sample embedding a large density of QDs. For the same pillar diameter, the connected pillar cavity fundamental mode presents a slightly lower energy, evidencing a lower optical confinement as the  field  partially penetrates  in the connected wires (figure \ref{connected1}.b.). Comparing the quality factor for connected versus isolated pillars for the same confinement (same mode energy), we find that while connected pillar cavities present slightly lower quality factors, the latter can still reach $Q=30000$. Such high quality factors show that connected pillar cavities could be used to reach the strong coupling regime. 
Concerning light extraction efficiency, the slightly reduced quality factor compared to isolated pillars may indicate some additional side losses, due to light guided in the wires. However, since the out coupling efficiency depends on $\eta_{\mathrm{top}}=\frac{\kappa_{\mathrm{top}}}{\kappa_{\mathrm{top}}+\kappa_{\mathrm{bottom}}+\kappa_{\mathrm{loss}}}$
, it can still be brought close to one by adjusting the parameters so that $\kappa_{\mathrm{top}} \gg  \kappa_{\mathrm{loss}},\kappa_{\mathrm{bottom}} $.\\

\begin{figure}[h!]\begin{center}\includegraphics[width=0.9\linewidth]{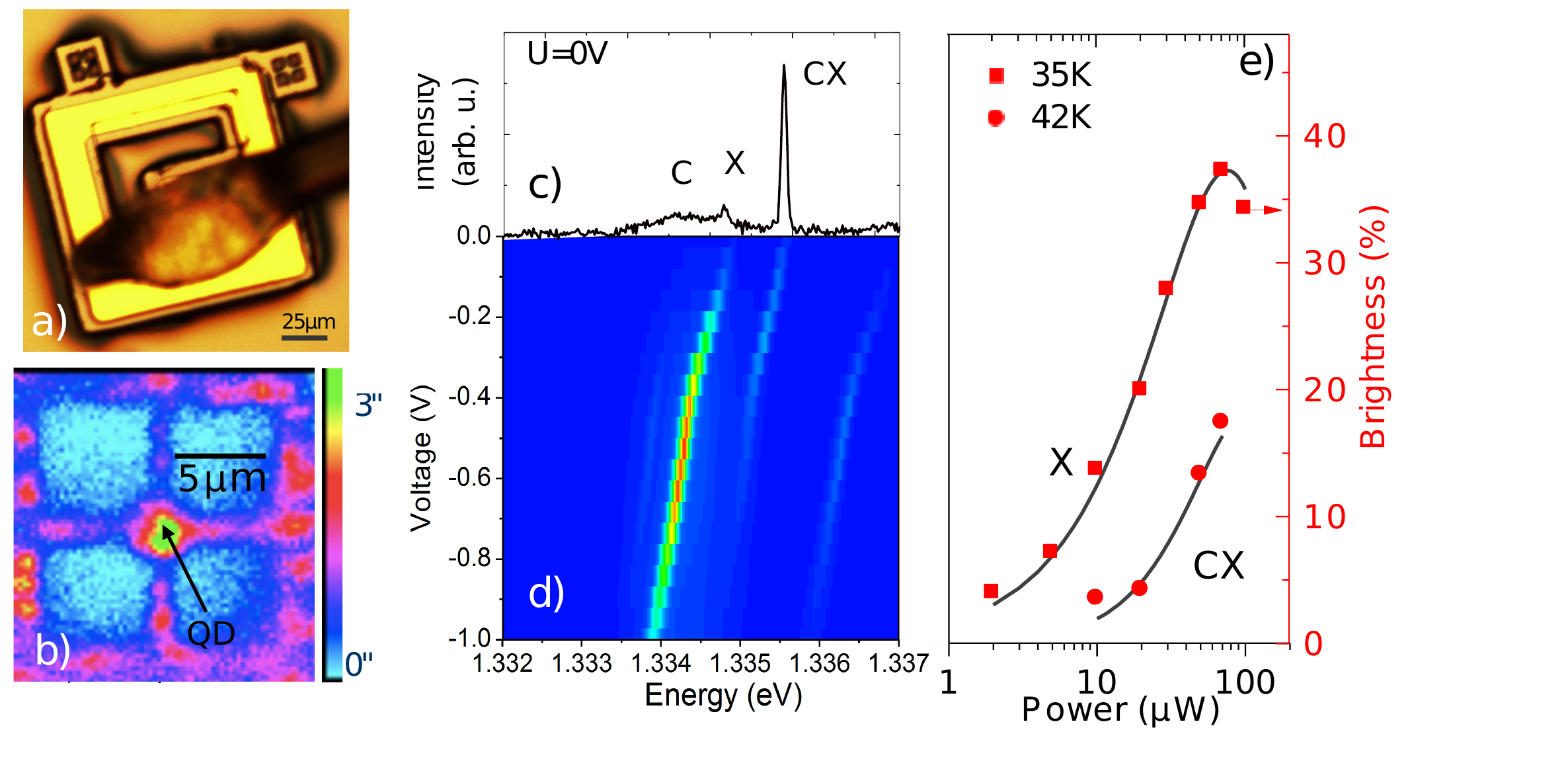}\caption{ {\bf Electrically tunable bright single photon source.} a: Microscope image of a device. The electrical contact and bonding are realized on a 100 $\mu m$ wide mesa. The pillars are connected to a 25 $\mu m$ wide frame overlapping the large mesa. b: Emission scan of the device (the sample is moved with respect the the excitation and confocal detection line). The emission is selected in a 5 nm wide spectral range around the cavity resonance. c: Emission spectrum without applied bias. The cavity (C), exciton (X) and charged exciton (X) lines are seen. d: Emission intensity as a function of applied bias. e: Brightness as a function of excitation power for the X line (squares) and CX line (circles). Lines are guides to the eyes.}\label{connected2}\end{center}\end{figure}

To deterministically insert a single QD in such a cavity, we have extended the in-situ lithography technique so as write any pattern in the resist, centered on a selected QD. This requires having a control on the absolute sample position with respect to the laser beam. Using a customized attocube confocal microscope, such a control was possible with a 10 nm accuracy, using high accuracy capacitive sensors. The  pillars, centered on a single QD, were connected to a 25~$\mu$m~$\times$~25~$\mu$m frame, the latter being connected to a $100 \ \mu m$ wide mesa. After resist development, metallic deposition and etching of the pillar structure, a second standard optical lithography step was used to define contact on the large mesa structure. Figure \ref{connected2}.a. presents an optical microscope image of a final device, where two connected pillars are visible on the right side. Figure \ref{connected2}.b. presents emission spectrum obtained under optical excitation when no bias is applied. The cavity line is slightly detuned from both the neutral X and charged CX exciton QD lines. The present structure embeds a p-i-n junction with a QD layer surrounded by barriers so as to allow the Stark tuning of the QD optical transition.  By increasing the voltage, the QD transition can be tuned over a 1.4 meV spectral range, throughout the cavity resonance (figure \ref{connected2}).  Finally, an emission map of the device  where the emission in a 5 nm spectral range around the cavity mode is selected is presented in figure \ref{connected2}.d.. The intense emission of the QD centered in the connected pillar is clearly evidenced, showing the Purcell enhanced extraction efficiency.\\

As for single pillars, a calibrated experimental setup is used to measure the  single photon source brightness (\ref{connected2}.e). In this experiment, the charge state of the QD was not well controlled, so that the QD under study is in the neutral and charged exciton state with 0.69 and 0.31 relative probabilities. When bringing either neutral or charged exciton lines in resonance with the cavity mode, the measured brightness reaches respectively $37 \pm 7\%$ and $17 \pm 6\%$. This corresponds to an extraction efficiency of $54\%$, limited here by the low Purcell factor of the source ($F_P=0.8\pm 0.08$, $\eta_{\mathrm{top}} \approx 1$, $\beta=0.44$). \\

In this first technological realization, we demonstrated the electrical tunability for a bright single photon source. With a different doping structure, the same technology can be used to control the charge state of the QD and build a deterministic spin-photon interface. Resonant spectroscopy are currently investigated in such structures. While the charge state of the QD in cavities has been shown to be sometime  unstable under resonant spectroscopy \cite{Reinhard2011}, preliminary tests indicate a significantly improved situation in gated structures.

\subsection{Implementation of an entangling CNOT gate}

To demonstrate the potential of QD based bright single photon sources for quantum information processing we have implemented an entangling controlled-NOT (C-NOT) gate \cite{gazzanoprl2013}. Indeed, a universal quantum computer can be  built with solely C-NOT gates and arbitrary local rotations, the latter being trivial in optics. A C-NOT gate flips the state of a target qubit depending on the state of a control qubit. Here the two qubits are single photons successively generated by a single QD-pillar based source with a brightness of $78\pm 7 \%$. The information is encoded on the polarization of the photons. \\

\begin{figure}[h!]\begin{center}\includegraphics[width=0.8\linewidth]{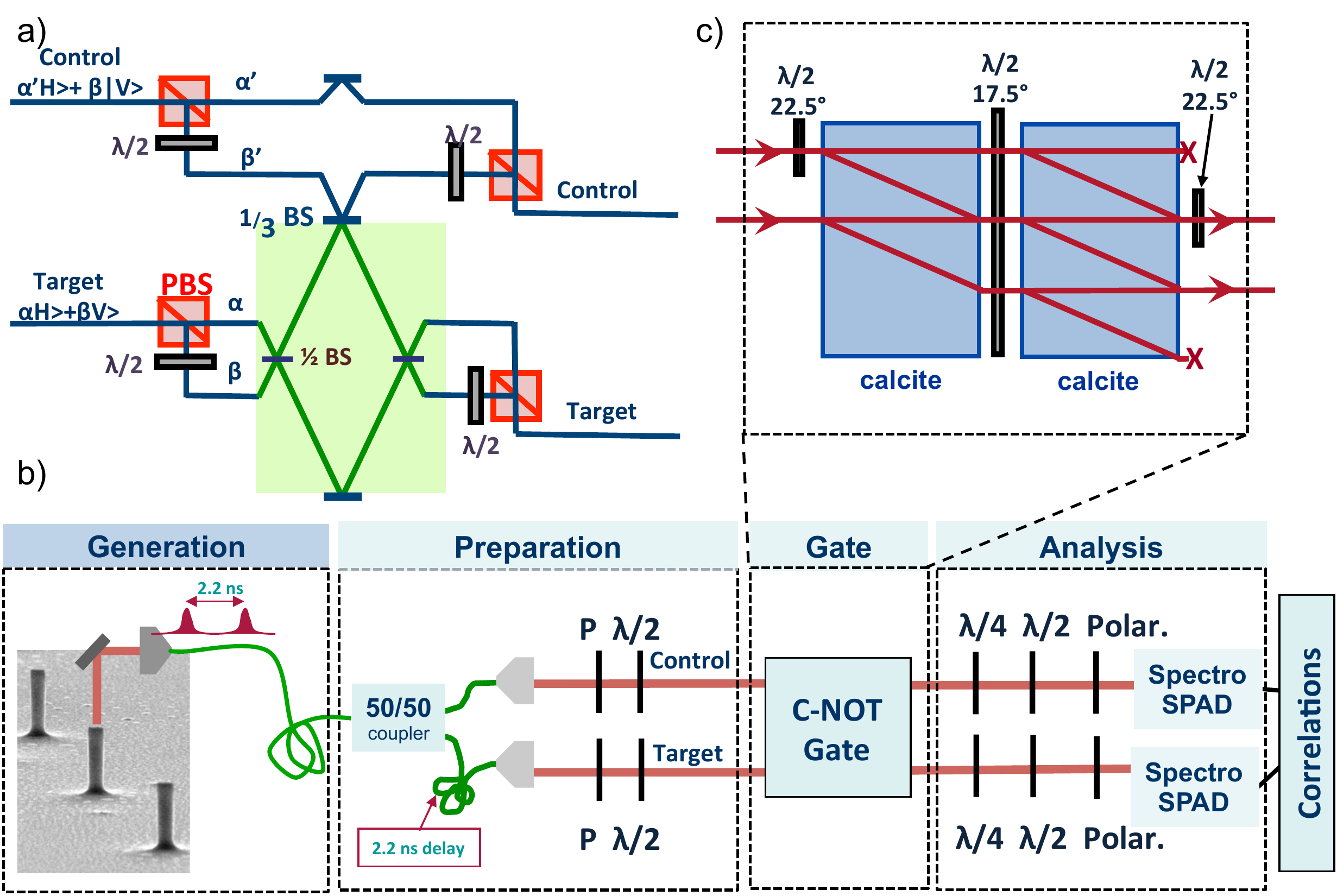}\caption{ {\bf Implementation of a C-NOT gate}. a: Schematic of a possible implementation for a C-NOT gate. b: Experimental setup used to entangle two photons generated by a QD based single photon source using a C-NOT gate.  c: Zoom on the C-NOT gate implemented using calcite crystals and wave plates. See text for details. }\label{CNOT1}\end{center}\end{figure}

Figure \ref{CNOT1}.a. illustrates a possible implementation of  an optical C-NOT gate . We first consider only the path concerning the target qubit, in a linear superposition of  $H$ and $V$ polarization $\ket{target}_{in}=\alpha \ket{H}+\beta \ket{V}$. This polarization encoding is tranformed into a path encoding using a polarizing beam splitter and a half-wave-plate. The two paths are then sent to the two input of a Mach-Zender interferometer (MZI). At the output of the interferometer, another half wave plate and polarizing beam splitter return from path  to polarization encoding. If the phase difference between the two arms of the MZI is $\pi$, the target qbit is flipped into the $\ket{target}_{out}=\alpha\ket{V}+\beta \ket{H}$. To implement a C-NOT gate, the MZI is set to a zero phase difference between the two arms, and the $\pi$ phase shift of one arm is induced by the controlled qubit. To do so, one arm of the MZI embeds a 1/3 beam splitter. The control qubit (upper part of \ref{CNOT1}.a.) is path encoded, one path being sent on the 1/3 beam splitter of the MZI. When the control and target qubit are indistinguishable, their quantum interference results in an effective $\pi$ phase shift between the two MZI arms. It can be shown that such a conceptually simple scheme acts as a quantum C-NOT gate on single photons. However, such an experimental scheme is hard implement because it requires stabilization of the optical paths. Here we use a simpler way to implement such a gate proposed in 2003 \cite{cnotdesign} and  is illustrated in Figure \ref{CNOT1}.b. It relies on two calcite crystals  implementing the path encoding and the interferometer and an internal half wave late implementing the 1/3 beam splitters. \\ 

Two photons  generated by the source with a delay of 2.3 ns are coupled to a single mode fiber. The coupling efficiency into the fiber is $82 \%$ as measured by comparing the count rate with and without the fiber coupling. The two photon are then non-deterministically split on a 50/50 fiber beam splitter and temporally overlapped using a 2.3 ns delay line on one arm of the fiber splitter. The two photons are then sent to the free space C-NOT gate setup, where waveplates before and after the gate allows controlling and analyzing  the  qubit polarization.\\

\begin{figure}[h!]\begin{center}\includegraphics[width=0.65\linewidth]{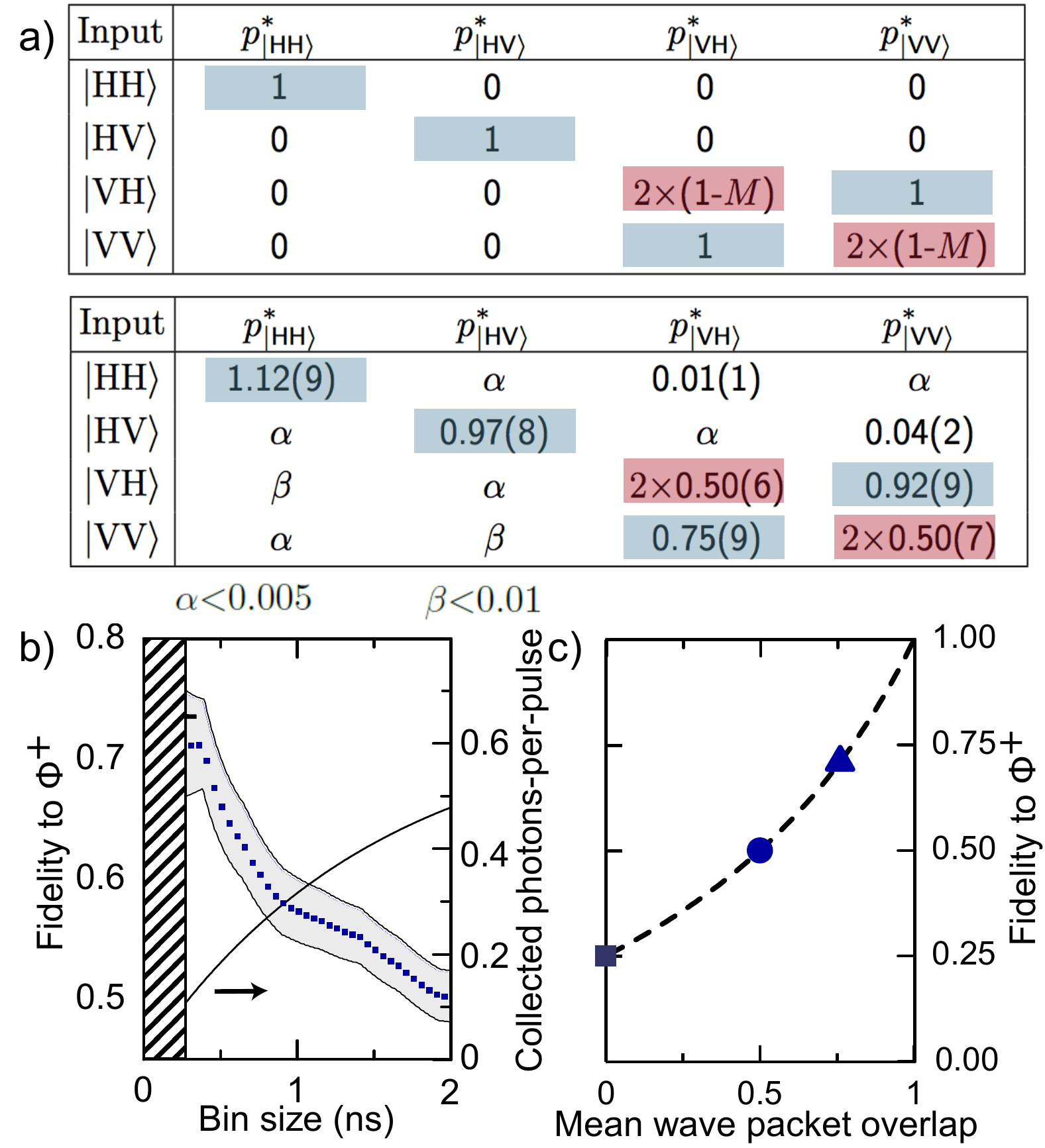}\caption{ {\bf Entangling capability of the gate} a: theoretical (top) and experimental (bottom) truth table of the gate. M is the mean wave packet overlap of the photons. b: Fidelity to the Bell state as a function of the bin size (left axis) and corresponding brightness (right axis). c: Fidelity to the Bell state as a function of the mean wave packet overlap. line: theoretical curve, symbols measured point for a brightness of 50 \% (circle) and 17\% (triangle) }\label{CNOT2}\end{center}\end{figure}

 The measured truth table of the gate in the linear $H$ and $V$ polarization basis is presented in figure \ref{CNOT2}.a. together with the calculated truth table for a mean wave packet overlap $M$ characterizing the indistinguishability of the two photons. In the ideal case, $M=1$, the target qubit is flipped from $H$ to $V$ (and vice versa) when the control qubit is set to $V$.  The observed truth table deviates from the ideal one because the indistinguishability of the photon is not ideal. The measurements are consistent with an experimental photon overlap of $M=0.5$. Note that this value is not the photon mean wave packet overlap as presented in figure \ref{indish} where the photon indistinguishability was deduced after correcting from the setup imperfection. Here, the raw photon wavepacket overlap is deduced from this measuments. The deduced value is consistent with ones reported earlier for a source  operated at a brightness of $75 \%$. The probability of obtaining the correct output averaged over four possible inputs is measured to 68.4\% for a maximal source brightness. Because the photon exhibits a better indistinguishability at short time delay (figure \ref{indishbin}), the probability of obtaining the correct output  increases to 73\% for a source brightness of 17\%. \\

To prove the entangling capability of the gate,  the control qubit is set to $\ket{D} = (\ket{V}{+}\ket{H})/\sqrt{2}$, and the target qubit to $\ket{H}$.  The output of an ideal gate is then the maximally entangled state $\Phi^{+} = (\ket{V,V}{+}\ket{H,H})/\sqrt{2}$. The fidelity of the two photon state generated experimentalled is deduced by measuring the polarization of the correlation in three polarization bases  bases~\cite{White07,measuringqbit}:
$$ E_{\alpha,\beta}=\frac{A_{\alpha,\alpha}+A_{\beta,\beta}-A_{\alpha,\beta}-A_{\beta,\alpha}}{A_{\alpha,\alpha}+A_{\beta,\beta}+A_{\alpha,\beta}+A_{\beta,\alpha}}$$ where $A_{\beta,\alpha}$ is the zero delay peak area in the correlation measurements for the output control photon detected in $\beta$ polarization and the output target photon in $\alpha$ polarization. The fidelity to the Bell state is  given by $F_{\Phi^{+}} = \left(1{+}E_{H,V}{+}E_{D,A}{-}E_{R,L}\right)/4$ where D,A refer to the diagonal and the anti-diagonal polarisation, and R and L to right and left  circular polarisations. \\

For  entanglement measurements,  the source brightness has been set to $I_{max}$=65 \% so as to benefit from a better degree of indistinguishability of the photons. The fidelity to the Bell state $F_{\Phi^{+}}$ is presented in \ref{CNOT2}.b. as a function of time bin, with the corresponding source brightness indicated on the right scale. For the maximum brightness, the fidelity to the Bell state is above the 0.5 limit for quantum correlations.  When reducing the time bin, the fidelity increases up to $0.710{\pm}0.036$. The theoretical fidelity to the Bell state is $F_{\Phi^{+}} = \frac{1+M}{2(2-M)}$ is plotted on  Figure \ref{CNOT2}.c. as a function of the mean wavepacket overlap, $M$. For maximum brightness, a fidelity of 0.5 correspond to M=0.5 (circle). For a time bin of 400 ps, the measured fidelity gets as high as 0.71, corresponding to mean wavepacket overlap larger than M=0.76 (triangle).

While we have reported the first implementation of an entangling C-NOT gate using a QD based single photon source, our study shows that a significant improvement of the indistinguishability is still needed to make QD based sources suitable for optical quantum computing. We discuss in section \ref{sec_future_challenges} ways to reach such a goal.
\section{Nonlinear optics with few-photon pulses}
\label{sec_few_photon_nonlin}

In this section, we now address a symmetric situation, where a QD in a cavity is studied to implement a single photon router.
\subsection{Motivations: photon blockade and photon routing}

\begin{figure}[h]
\centering
\includegraphics[width=10cm]{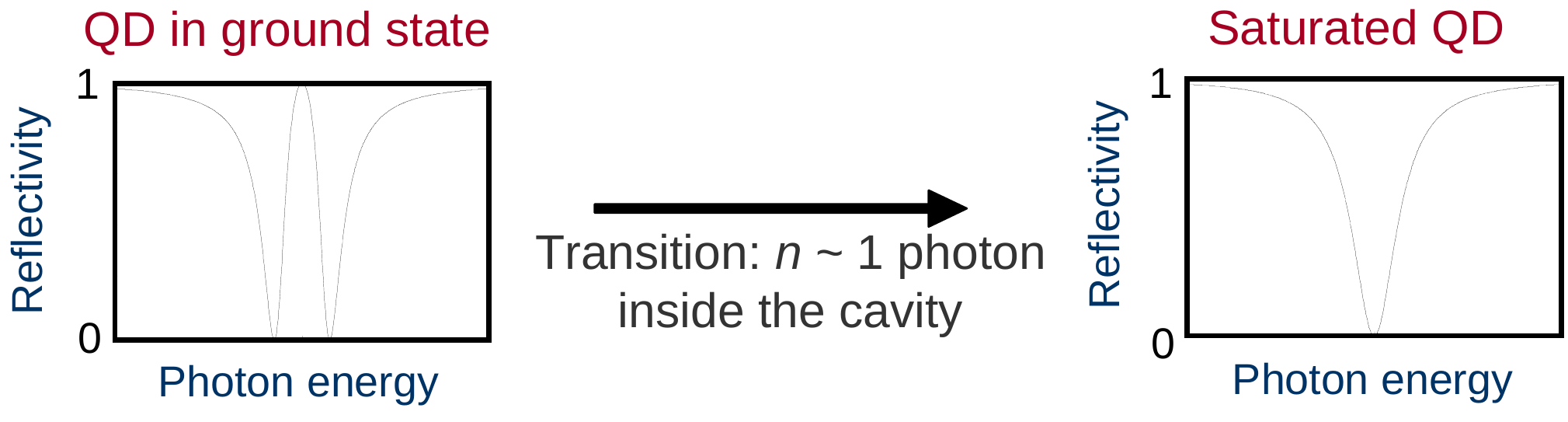}
\caption{Theoretical reflectivity spectra for a strongly-coupled cavity-QED device, in the two limiting cases of a quantum dot in the ground state and of a saturated quantum dot transition.}
\label{fig_two_behaviors_th}
\end{figure}

A two-level system is, by nature, a strongly nonlinear system: it may interact with a first photon but, once the two-level transition is saturated, it will not interact with a second one. In the absence of a cavity structure, taking advantage of this with a quantum dot is very inefficient: most of the photons incident on the quantum dot will not interact with it. It is much more useful to use a QD in a cavity-QED device, as in such a case the optical properties of the system can dramatically depend on the QD state. As an example, Figure \ref{fig_two_behaviors_th} describes the theoretical reflectivity spectrum of a strongly-coupled device having both high cooperativity $C\gg1$ (see definition in Eq. \ref{eq_coop}) and a top-mirror output-coupling adjusted to $\eta_{\mathrm{top}}=50\%$ (for example with a symmetrical design where $\kappa_{\mathrm{top}}=\kappa_{\mathrm{bottom}}=\frac{\kappa}{2}$, and $\kappa_{\mathrm{loss}}=0$). When the quantum dot is in its ground state, the reflectivity spectrum presents two dips associated to the two eigenstates of the system, separated by the Rabi splitting $2g$. When the QD transition is saturated, on the contrary, the reflectivity spectrum presents a single Lorentzian dip associated to the cavity mode resonance. A continuous transition between these two behaviors can be obtained when, increasing the incident power, the average number of photons in the cavity approaches unity. This nonlinear effect has been named ``giant optical nonlinearity'' due to this extremely low photon number nonlinearity threshold \cite{Auffeves-Garnier2007}.\\

How can such an effect be exploited for practical applications? The main idea is that the transmission/reflection probability for a second photon will be modified if a first one has already been incident on the device. One can use this nonlinearity to engineer quantum light from a classical laser beam: this is the \emph{photon-blockade effect}  ensuring, for instance, that no more than one photon at the same time will be transmitted by the cavity \cite{Birnbaum2005}. Another important device for quantum applications would be a \emph{single-photon router}: a device so nonlinear that, if two photons are simultaneously incident on it, the first one will get transmitted and the second one reflected \cite{Chang2007}. This would constitute a major breakthgrough for quantum information and communication. Indeed, contrary to the coalescence of indistinguishable photons on a beamsplitter cube (see previous section), it would allow the engineering of a \emph{deterministic} interaction between two photons, mediated by the cavity-QED device. \\

Towards this final objective of single-photon routing, several realizations have already been obtained in various types of microcavity systems. Recently, resonant spectroscopy on coupled QD-cavity devices, in the form of photonic crystals \cite{Englund2007,Reinhard2011,Englund2012,Bose2012,Volz2012} or microdisks \cite{Srinivasan2007,Srinivasan2008}, has demonstrated giant optical nonlinearity and fast optical switching. These works all concluded that optical nonlinearity is obtained when close to unity photon numbers are reached \emph{inside} the cavity. However, hundreds of incident photons were required to obtain a single intracavity photon. For future quantum applications,  distinguishing between the intracavity photon number and the number of \emph{incident photons per pulse} is crucial. An optical nonlinearity behavior at the level of one to two incident photons per pulse is needed: as described below, the current record is an optical nonlinear threshold at 8 photons per pulse recently achieved using a QD-micropillar device \cite{Loo2012}.

\subsection{Observation of nonlinearities at the few-photon scale}

\begin{figure}[h]
\centering
\includegraphics[width=9cm]{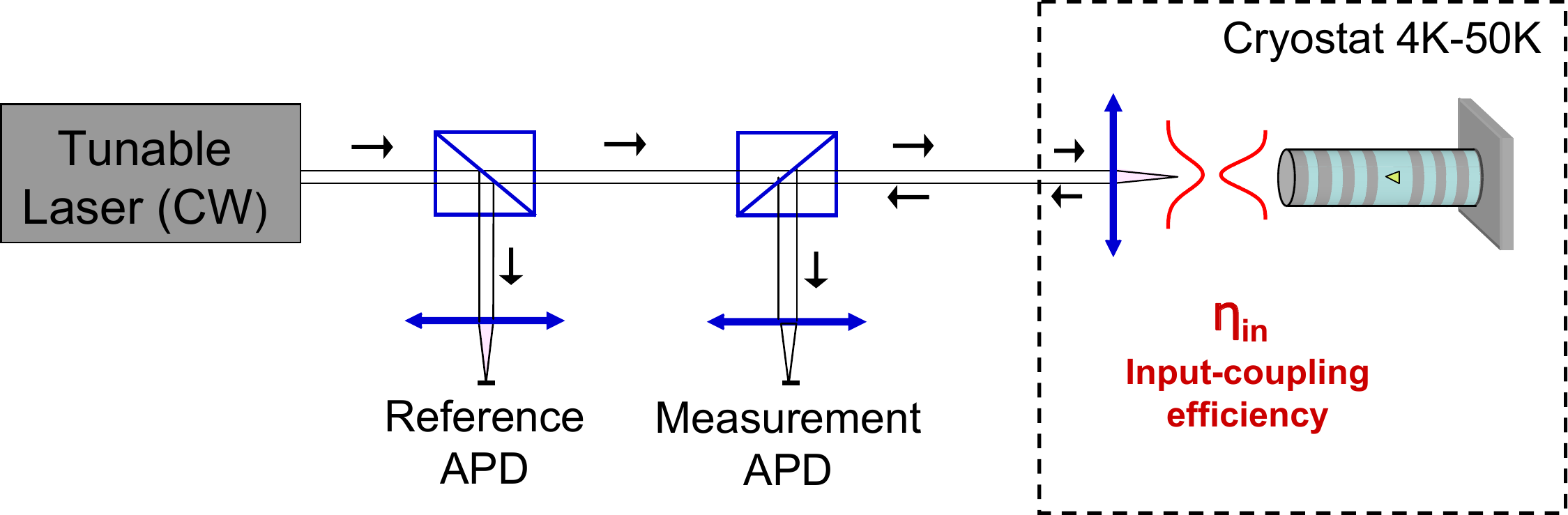}
\caption{Simplified sketch of a resonant excitation setup for reflection spectroscopy measurements.}
\label{fig_refl_setup}
\end{figure}

The results described here have been obtained with a QD-pillar device which is in the strong-coupling regime thanks to a very high quality factor Q=29000, for a $2.1 \mathrm{\mu m}$ diameter. This could be obtained using the in-situ lithography technique on a sample where the bottom and top Bragg mirrors are constituted by 36 and 32 pairs, respectively, so that they have equal reflectivities and thus equal damping rates $\kappa_{\mathrm{top}}=\kappa_{\mathrm{bottom}}$. A simplified sketch of a resonant excitation setup, allowing the measurement of a device reflectivity spectrum with  high spectral resolution, is displayed in Fig. \ref{fig_refl_setup}. The sample is placed inside a helium vapor cryostat, altogether with a focusing lens, the sample position being controlled with nanopositioners inside the cryostat. A CW or pulsed laser is injected into and reflected from the micropillar with a finely tunable photon energy $\hbar \omega$. Non-polarizing beamsplitters are used to split the incident and reflected beams: the incident power is measured with a first avalanche photodiode, a second one being used to measure the reflected power. The input-coupling efficiency $\eta_\mathrm{in}$ is optimized thanks to a careful optical alignment. \\

\begin{figure}[t]
\centering
\includegraphics[width=10cm]{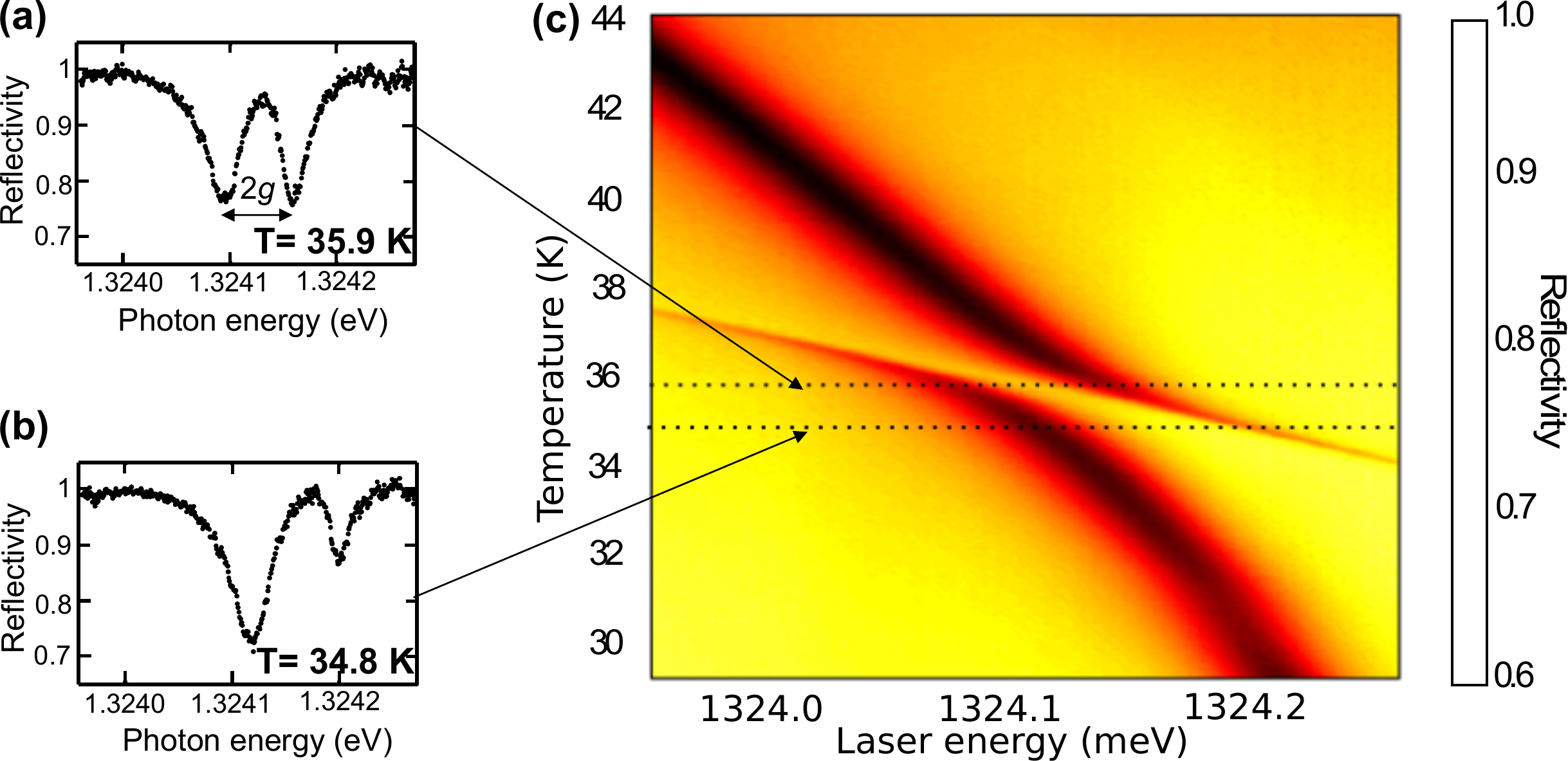}
\caption{\textbf{(a)} and \textbf{(b)} Reflectivity spectra at low incident power, for different temperatures $T=35.9K$ and $T=34.8K$. \textbf{(c)} Reflectivity map as a function of device temperature and laser photon energy.}
\label{fig_strong_coupling}
\end{figure}

Figure~\ref{fig_strong_coupling}a presents the reflectivity spectrum measured, at low incident power, at the resonance temperature $T=35.9K$: The system is in a pronounced strong-coupling regime, the two reflectivity dips being associated with the exciton-photon eigenstates of the system, having equal excitonic and photonic parts. Figure~\ref{fig_strong_coupling}b presents a reflectivity spectrum measured at a different $T=34.8K$ where the asymmetrical shape arises from the unequal excitonic and photonic parts for the exciton-photon eigenstates. A a final characterization of the device behavior at low power, Figure~\ref{fig_strong_coupling}c shows an experimental map of the reflectivity measured as a function of both temperature and photon energy $\hbar \omega$, where the darker areas correspond to lower reflectivities. The low-reflectivity regions directly evidence the temperature dependence of the two coupled exciton-photon eigenstates, and their anticrossing when the device temperature is tuned \cite{Loo2010}. \\

\begin{figure}[h]
\centering
\includegraphics[width=9cm]{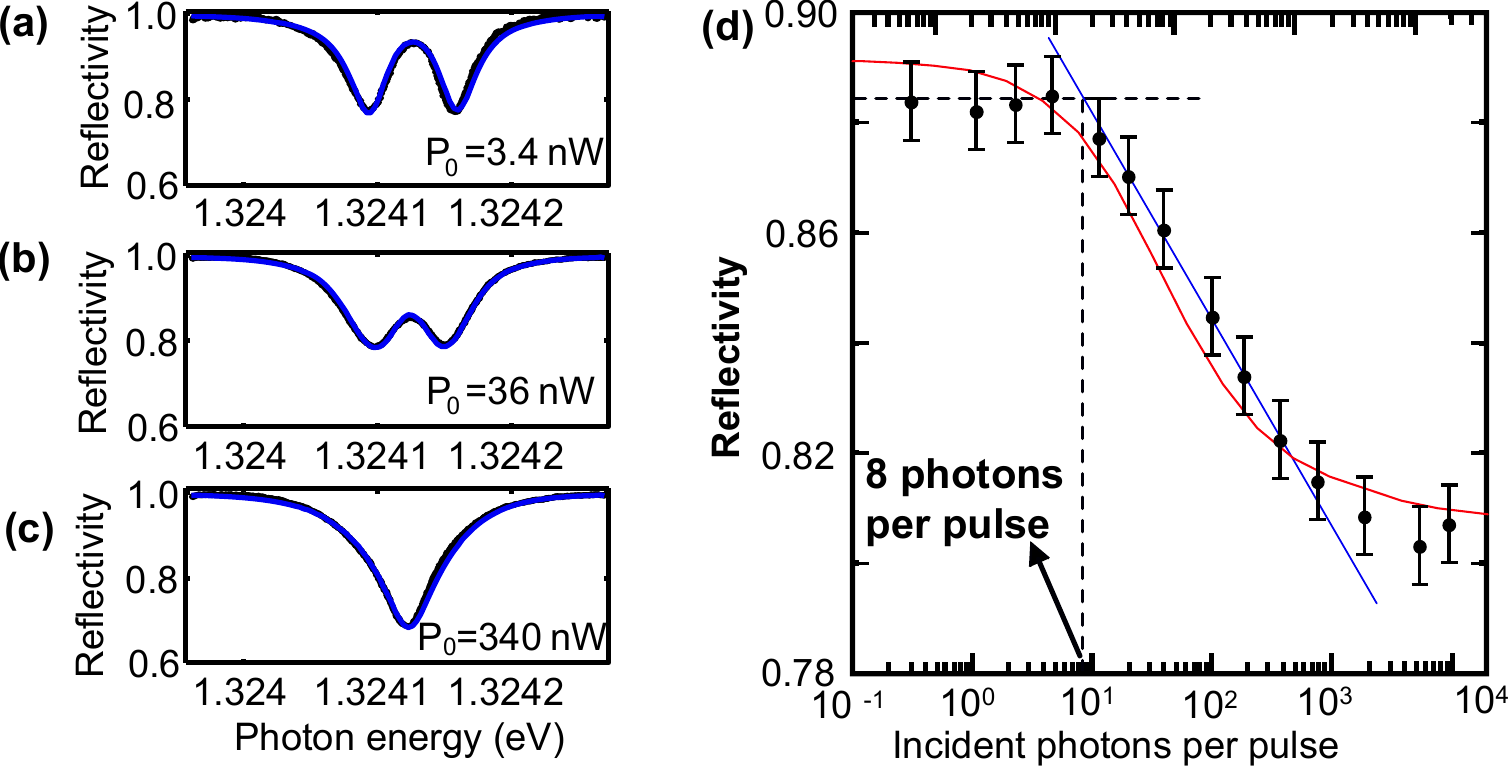}
\caption{\textbf{(a)} to \textbf{(c)} Reflectivity spectra for various incident powers, illustrating the nonlinear transition under CW excitation. \textbf{(d)} Pulsed excitation: reflectivity measurement displaying a record nonlinearity threshold at 8 incident photons per pulse.}
\label{fig_exp_nonlin}
\end{figure}

Figures \ref{fig_exp_nonlin}a to \ref{fig_exp_nonlin}c illustrate the nonlinear behavior of this device under CW excitation with a varying pump power \cite{Loo2012}. A transition is observed from the low-power regime (two reflectivity dips) to the high-power regime (single reflectivity dip). Fitting these experimental data allows determining the figures of merit of our cavity-QED device: a good cooperativity ($C=2.5$) and a near-unity input-coupling ($\eta_\mathrm{in}=95\%$), but a relatively low top-mirror output-coupling  ($\eta_\mathrm{top}=8\%$ instead of $50\%$ for an ideal device). The high cooperativity is related to the very good contrast, at low power, between the two reflectivity dips. The quite low top-mirror output coupling $\eta_\mathrm{top}=8\%$ is the reason why the minimal reflectivity is not zero, contrary to the ideal situation described in Fig. \ref{fig_two_behaviors_th}.\\

Finally, Figure \ref{fig_exp_nonlin}d reports a reflectivity measurement under pulsed excitation, with an optimized optical pulse whose spectral width matches that of the cavity mode resonance. The device reflectivity is plotted as a function of $N$, the number of incident photons in each pulse. As can be seen, a nonlinearity threshold at 8 incident photons per pulse is obtained \cite{Loo2012}. This constitutes a record value which became achievable thanks to the near-unity input coupling efficiency in our micropillar: the previous record was a threshold at 80 photons per pulse with a  photonic crystal cavity \cite{Bose2012}. Figure \ref{fig_exp_nonlin}d also shows that the experimental data fit with the predictions of cavity-QED, using the same parameters as used with the CW experiment ($C=2.5$, $\eta_\mathrm{in}=95\%$, $\eta_\mathrm{top}=8\%$).

\subsection{Device optimization: towards a single-photon router?}

Looking at the device figures of merit, it is clear that the improvement margin lies in the top-mirror out-coupling, which should be brought closer to the $50\%$ ideal value. This requires decreasing the loss damping rate and/or increasing the mirror damping rates, so that $\kappa_\mathrm{loss}\ll\kappa_\mathrm{top}=\kappa_\mathrm{bottom}$. Reducing the sidewall losses by increasing the pillar diameter, or increasing the mirror damping rate by decreasing the number of layers in the Bragg mirrors, is a first way to do so. This, however, would require a careful optimization as it could also degrade the device cooperativity. Another possibility is to use adiabatic cavities, following Lermer et al \cite{adia}, which allows decreasing $\kappa_\mathrm{loss}$ without increasing the pillar diameter; it provides a way to increase both the top-mirror output coupling $\eta_\mathrm{top}$ and the cooperativity $C$ (through the decrease of the total damping rate $\kappa=\kappa_\mathrm{top}+\kappa_\mathrm{bottom}+\kappa_\mathrm{loss}$). \\

\begin{figure}[h]
\centering
\includegraphics[width=9cm]{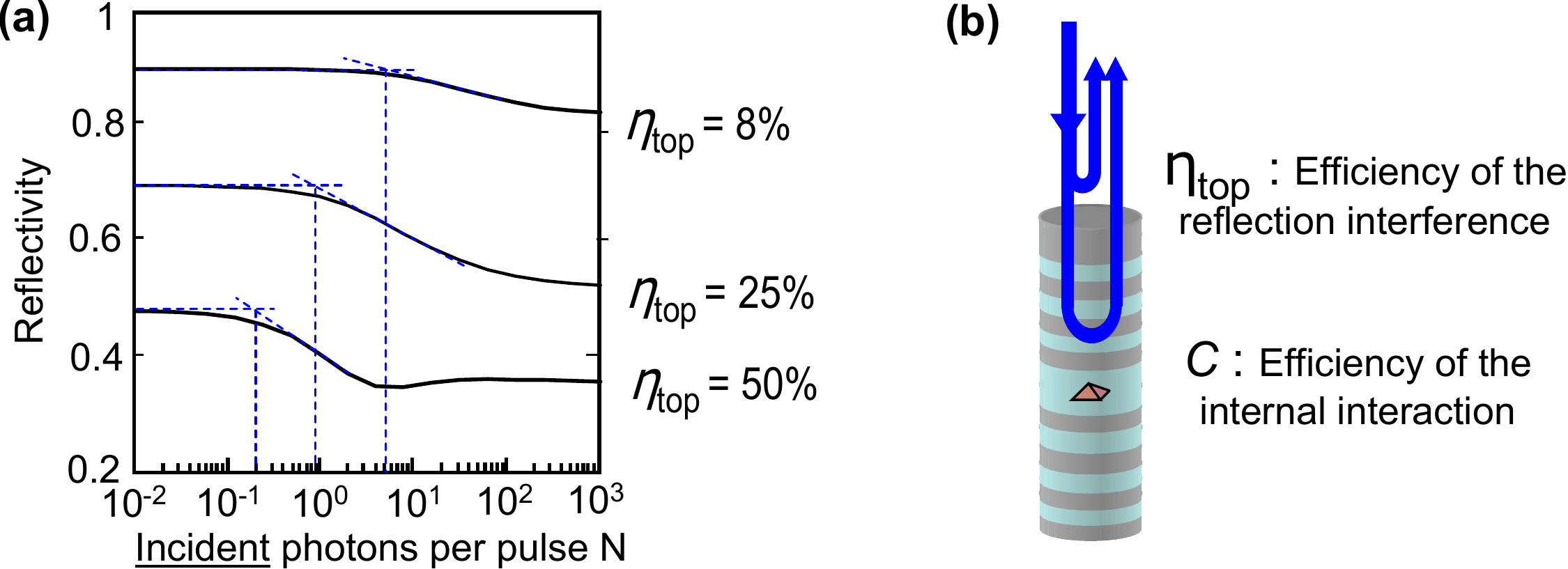}
\caption{\textbf{(a)} Predicted nonlinear behaviors under pulse excitation, for various top-mirror output couplings $\eta_\mathrm{top}$. \textbf{(b)} Sketch of the interference, in the reflected beam, between the directly-reflected field and the field which has interacted with the quantum dot and been reextractedo out through the top-mirror.}
\label{fig_nonlin_pulse_th}
\end{figure}

To illustrate the impact that such an improvement would have on the nonlinear device, Figure \ref{fig_nonlin_pulse_th}a displays the theoretical device reflectivity as a function of $N$ for increasing output couplings $\eta_\mathrm{top}$. We find that a factor six increase in $\eta_\mathrm{top}$ decreases the expected threshold by a factor 30 \cite{Loo2012}. As sketched in Figure \ref{fig_nonlin_pulse_th}b, this is explained by the fact that the reflected beam results from the interference between two fields: a directly-reflected field and a field that has been injected into the cavity (input coupling $\eta_\mathrm{in}$), has interacted with the quantum dot (cooperativity $C$), and has been re-extracted out through the top mirror (output coupling $\eta_\mathrm{top}$). Increasing $\eta_\mathrm{top}$ is thus crucial to significantly increase the strength of this interference.\\

Furthermore, one finds that a nonlinearity threshold lower than $1$ can be obtained with an optimized top-mirror output coupling, so that for $N=1$ incident photon per pulse the system will be precisely in the region of highest nonlinearity. This paves the way toward the realization of single-photon routers and quantum logic gates operating with single-photon incident pulses. However, we must point out that the calculations presented here are performed with attenuated coherent pulses, rather than with true one-photon or two-photon pulses. Actually the road towards \emph{deterministic} single-photon routers (which transmit a first photon with $100\%$ probability and reflect a second one with $100\%$ probability) is still long: it will not only require technological improvements but also experimental schemes a bit more complex than the two-level system nonlinearity \cite{Rosenblum2011}.

\subsection{Resonant excitation: application to fast optical nanosensing}

A quantum-dot strongly-coupled to a cavity mode is an extremely sensitive device whose optical properties can be controlled in several other ways. For instance, it can be sensitive even to very small electrostatic fluctuations, like those induced by the motion of carriers in the vicinity of the quantum dot. Indeed, a slight modification of the QD electrostatic environment can induce a small variation of the QD optical transition frequency $\omega_\mathrm{QD}$. This variation, in turn, can strongly change the device reflectivity and be readily detected with an appropriate resonant excitation setup.\\
\begin{figure}[h]
\centering
\includegraphics[width=8cm]{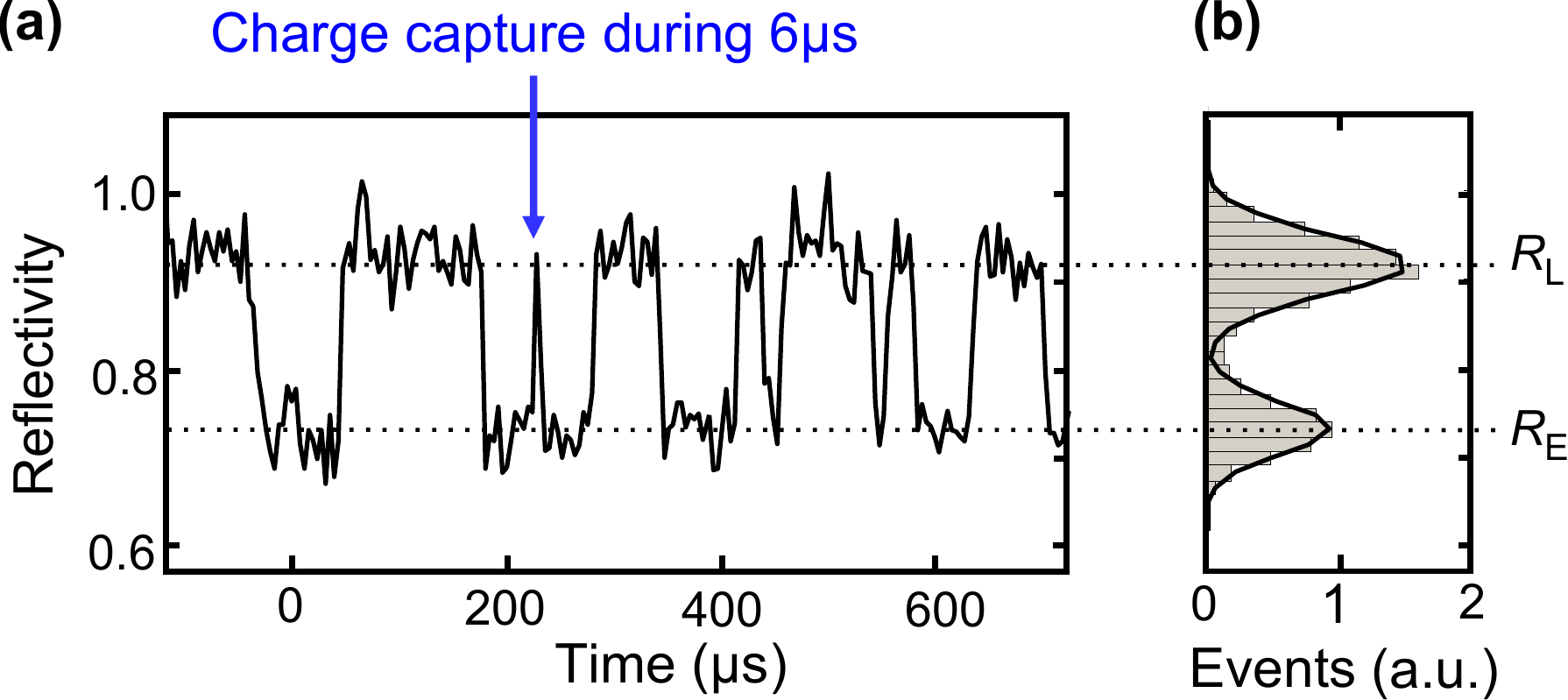}
\caption{\textbf{(a)} Real-time reflectivity measurement, monitoring the capture and subsequent release of individual carriers by a single material defect. \textbf{(b)} Histogram showing that the reflectivity randomly jumps between two values, $R_L$ and $R_E$, respectively corresponding to a loaded or to an empty trap.}
\label{fig_monitoring}
\end{figure}
Using a strongly-coupled device very similar to the one used for the optical nonlinearity measurements, it has recently been possible to monitor in real-time single quantum events, corresponding to a carrier being captured and then released by a material defect. The experiment could be performed at the microsecond time scale \cite{Arnold2014}: this measurement rate is five orders of magnitude faster than for previous optics experiments of single-charge sensing, because of the close to shot-noise-limited detection setup and of the enhanced light-matter interaction. Figure \ref{fig_monitoring}a displays a typical real-time reflectivity signal illustrating the monitoring of single-charge jumps between two states (loaded/empty material trap). The vertical arrow, for example, indicates an event where a single charge has been captured by the material trap and then released 6 $\mathrm{\mu}s$ later. The clear distinction between the loaded and empty states is also illustrated in the reflectivity histogram of Fig. \ref{fig_monitoring}b: the overlap between the two distributions is small enough to allow identifying the system state, at any time, with a less than $0.2\%$ error probability. This powerful resonant excitation technique can also be applied to the real-time monitoring of other rapid quantum events such as the spin flips of a single electron or hole resident in a charged quantum dot: such an experiment would constitute the building block a spin-photon interface.

\section{Future challenges}
\label{sec_future_challenges}
 The recent advances in QD based  technologies make them very good candidates for fabricating the next generation of single photon sources used in optical quantum computing. While the source brightness has reached very high values, the indistinguishability of bright single photon sources needs further improvements. In this matter, controlling the electrostatic environment of the QD appears as a critical step. While such a control is more difficult to obtain in photonic structures like micropillars and nanowires where the QD is close to etched surfaces,  preliminary results on connected pillar devices indicate that such a control is within reach. \\

A very bright source of highly indistinguishable photons would have immediate applications in optical quantum computing, where a large number of photons successively emitted  by the same source would be temporally overlapped using appropriate delay lines. Some comment should however been made here: in most experiments, the indistinguishabilities of the successively emitted photon is tested with a limited time delay between the two photons (typically several nanoseconds). Indistinguishabilities on long time scales has not been tested yet. We note however that a recent study of the charge noise on gated QDs shows Fourier transform limited linewidth on a time scale as long as $20 \ \mu s$ \cite{warburton2013}. \\

In the long term, demonstrating the possibility to use several sources is crucial for the scalibility of a QD based quantum network. Quantum interferences between remote QD sources have  first been demonstrated in 2010 \cite{patel2010,solomon2010}. Impressive progresses have been reported recently using pure resonant excitation \cite{gaoteleportation}. Similar experiments are currently conducted using deterministically coupled QD-pillar bright sources. To that end, QD with similar optical transitions energies are inserted in pillars presenting the same diameter. Preliminary results show that the Purcell effect relaxes the requirement on the spectral matching between the two sources. It can also enable quantum interferences with a single photon source presenting a very low degree of indistinguishability for successively emitted photons.\\

\begin{figure}[h]
\centering
\includegraphics[width=10cm]{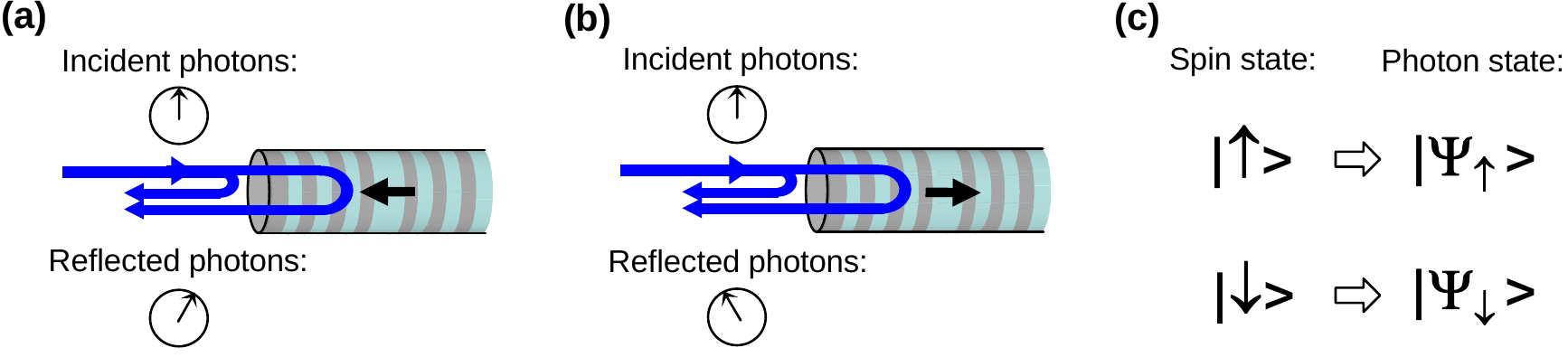}
\caption{\textbf{(a)} and \textbf{(b)}  Illustration of spin-dependent polarization rotation, induced by a single quantum dot spin. \textbf{(c)} Mapping from a spin state to a photon polarization state.}
\label{fig_spin_photon_interface}
\end{figure}

In section \ref{sec_few_photon_nonlin}, we saw that single photon switches based on a single QD coupled to a cavity are within reach, with a reasonable improvement of the current technology. While such optical non linearities are highly desirable, they present a limitation for applications: the photons must overlap temporally within the cavity lifetime. A  promising approach to engineer an interaction between \emph{delayed} photons is to insert a spin in a cavity: this requires a charged quantum dot, containing a resident carrier whose spin state can be used as an optically-accessible quantum memory. The basic concept at the heart of a spin-photon interface  is illustrated in Fig. \ref{fig_spin_photon_interface}a and \ref{fig_spin_photon_interface}b: if an input beam with a given polarization is injected into a QD-micropillar device, the reflected output beam will be rotated clockwise or counter-clockwise, depending on the spin state \cite{faraday1, faraday2}. In quantum words (see Fig. \ref{fig_spin_photon_interface}c), the reflected photons will be in the polarization state $\ket{\Psi_{\uparrow} }$ if the QD spin is in state $\ket{\uparrow }$, and in the polarization state $\ket{\Psi_{\downarrow} }$ if the spin is in state $\ket{\downarrow }$. This is the well known Faraday/Kerr rotation effect, a phenomenon widely used to optically characterize magnetic materials, but applied here to quantum physics with a single spin. \\

As with the previous experiments based on resonant excitation, the device figures of merit which will govern the efficiency of the polarization rotation are the cooperativity $C$ and the top-mirror output-coupling $\eta_\mathrm{top}$. Analytical calculations show that, for realistic values of $C$ and $\eta_\mathrm{top}$, the two possible output polarization states can be made orthogonal: $<\Psi_{\uparrow}\ket{\Psi_{\downarrow} }=0$. Such a configuration provides the possibility to reach a maximal entanglement between the state of the spin qubit and the polarization state of the output photon. Let us suppose that, before the interaction with a photon, the spin is first prepared in a coherent superposition  $\frac{1}{\sqrt{2}}\left( \ket{\uparrow }+ \ket{\downarrow }\right)$. Then, after interaction with an incident photon, the bipartite spin-photon system will end up in a maximally-entangled state of the form $\frac{1}{\sqrt{2}}\left( \ket{\Psi_{\uparrow} }\otimes\ket{\uparrow } + \ket{\Psi_{\downarrow} }\otimes\ket{\downarrow }\right)$. \\

In contrast to the recent spin-photon entanglement demonstrations \cite{spinphotonatac,spinphotonyoshie}, the interaction of a photon with such devices would allow the entanglement between a spin and a photon generated by an external source. Such a situation has been theoretically predicted to open new paradigms in quantum optics like delayed photon entanglement \cite{spin1}, deterministic logic gates \cite{spin2}  or fault-tolerant quantum computing \cite{spin3}. Recent measurements show that deterministically inserting a single spin in a pillar cavity indeed allows obtaining a rotation of the polarization by few degrees depending on the spin state. \\

Beyond the potential for quantum information processing, QD deterministically coupled to pillar cavities also opens the possibility to explore cavity quantum electrodynamics in a regime rarely explored by the atomic community, namely the broad emitter limit. Indeed, in the founding work by Purcell as well as for all experimental realizations with real atoms, the emitter presents a monochromatic spectrum with respect to the cavity linewidth. With solid state emitters, broadening induced by the environment give rises to new phenomena. We recently demonstrated that phonon assisted emission lead to cavity pulling phenomena for a single QD coupled to a cavity with moderate quality factor \cite{pulling}. Recent developments show that  phonon assisted Purcell effect can be used to obtain bright single photon sources, where strong coupling to the environment provides a built-in spectral tuning of the QD emission to the cavity resonance.

\section{Acknowledgements} %
\begin{acknowledgement}
The authors acknowledge their coworkers who have made all these results possible: Aristide Lemaitre, Isabelle Sagnes, Paul Voisin, Olivier Krebs, Adrien Dousse, Olivier Gazzano, Jan Suffczynski, Steffen Michaelis de Vasconcellos, Anna Nowak, Simone Luca Portalupi, Valérian Giesz, Niccolo Somaschi, Chirstophe Arnold, Vivien Loo, Justin Demory, Marcelo de Almeida, Andrew White and Alexia Auffeves. This work was partially supported by the French ANR DELIGHT, ANR MIND, ANR CAFE, ANR QDOM, the ERC starting grant 277885 QD-CQED, the CHISTERA project SSQN, the French Labex NANOSACLAY, and the RENATECH network.
\end{acknowledgement}

\section{References}

\bibliographystyle{SpringerPhysMWM} 
\bibliography{Senellart_chapter.bib}

\printindex
\end{document}